\def\spose#1{\hbox to 0pt{#1\hss}}
\def\approxlt{\mathrel{\spose{\lower 3pt\hbox{$\sim$}}
	\raise 2.0pt\hbox{$$<$$}}}
\def\approxgt{\mathrel{\spose{\lower 3pt\hbox{$\sim$}}
	\raise 2.0pt\hbox{$>$}}}

\def\multleft#1{\hbox to size{\vbox {\halign {\lft{##}\cr #1}}\hfill}\par}
\def\multright#1{\hbox to size{\vbox {\halign {\rt{##}\cr #1}}\hfill}\par}

\def\today{\ifcase\month\or January\or February\or March\or April\or May\or
      June\or July\or August\or September\or October\or November\or December\fi
      \space\number\day, \number\year}
\def\$<${\thinspace}
\def\s{\hbox{\phantom{5}}}	

\def\boxit#1{\vbox{\hrule\hbox{\vrule\kern3pt\vbox{\kern3pt
          #1 \kern3pt}\kern3pt\vrule}\hrule}}

\def\cm{{\rm\thinspace cm}}

\def\erg{{\rm\thinspace erg}}

\def\km{{\rm\thinspace km}}

\def\Msun{\hbox{$\rm\thinspace M_{\odot}$}}
\def\pc{{\rm\thinspace pc}}

\def\s{{\rm\thinspace s}}
\def\yr{{\rm\thinspace yr}}

\def\ergpcmsqps{\hbox{$\erg\cm^{-2}\s^{-1}\,$}}

\def\ergps{\hbox{$\erg\s^{-1}\,$}}

\def\kmps{\hbox{$\km\s^{-1}\,$}}

\def\Msunppc{\hbox{$\Msun\pc^{-3}\,$}}

\def\Msunpyr{\hbox{$\Msun\yr^{-1}\,$}}

\documentstyle[psfig]{mn}
\begin{document}
\hsize=6truein


\title[A {\it ROSAT} study of the cores of clusters of galaxies - I]
{A {\it ROSAT} study of the cores of clusters of galaxies - I: Cooling
flows in an X-ray flux-limited sample}

\author[C.B. Peres et al.]
{\parbox[]{6.in} {C.B. Peres, A.C. Fabian, A.C. Edge, S.W. Allen, 
R.M. Johnstone, and D.A. White \\
\footnotesize
Institute of Astronomy, Madingley Road, Cambridge CB3 0HA \\ 
}}                                            
\maketitle
%
%

\begin{abstract}  
This is the first part of a study of the  detailed 
X-ray properties of the cores of nearby clusters. We have used the 
flux-limited sample of 55 clusters of Edge et al. (1990) and  archival and 
proprietary data from the {\it ROSAT} observatory. 
In this paper an  X-ray spatial analysis based on the 
surface-brightness-deprojection technique is applied to the clusters in the 
sample with the aim of studying their cooling flow 
properties. We determine the fraction of cooling flows in this sample 
to be 70-90 percent and estimate the contribution of the 
flow region to the cluster X-ray luminosity. We show that the luminosity within a strong 
cooling flow can account for up to 70 percent of a cluster X-ray bolometric 
luminosity. Our analysis indicates that about 40 percent of the clusters 
in the sample have flows depositing more than 100 \Msunpyr throughout 
the cooling region and that these possibly have been undisturbed for many 
Gyr, confirming that 
cooling flows are the natural state of cluster cores. 
New cooling flows in the  sample are presented and 
previously ambiguous ones are clarified.
We have constructed a catalogue of 
some intracluster medium properties for the clusters in this sample.
The profiles of the mass deposited from cooling flows are analyzed and 
evidence presented for the existence of breaks in some of the 
profiles. Comparison is made to recent
optical and radio data. We cross-correlate our sample with the 
Green Bank, NVSS and FIRST surveys and to the volume-limited sample of 
brightest cluster galaxies  presented by Lauer and Postman (1994).
Although weak trends exist, no strong correlation between optical 
magnitude or radio power of the brightest cluster galaxy and the 
strength of the flow is found. 

\end{abstract}


\section{INTRODUCTION} 
\label{intro}

The X-rays emitted from clusters of galaxies represent a loss of energy 
of the intracluster gas. Thermal
bremsstrahlung and line emission depend on the square 
of the gas density, which rises towards the cluster centre.
As the density rises the loss rate increases and the resultant
cooling timescale, t$_{\rm cool}$ decreases. 
Within the cooling radius ($r_{\rm cool}$), the 
cooling time becomes smaller than the age of the cluster 
($t_{\rm cool}$ $<$ $t_{\rm age} \approx$ {\it H}$_0^{-1}$) and, under
gravity and thermal pressure, the  gas cools and flows 
inward to maintain pressure equilibrium. 

This simple picture describes the physics of a homogeneous cooling
flow. In a more realistic situation initial inhomogeneities in the
intracluster medium (ICM) lead to the establishment of a multiphase
atmosphere which cools over a wide range of radii depositing cool
material throughout the central $\sim$200 kpc of a cluster of galaxies. 
(For reviews on cooling flows, see Fabian 1994, 
and Fabian, Nulsen \& Canizares 1984).  

Although  the mass cooling out of the X-ray band  represents only 
a negligible amount when compared to the total hot phase of the ICM,
it is comparable to the mass of a brightest cluster galaxy (BCG) for flows which are 
undisturbed for about a Hubble time.
Therefore understanding the physical processes of a cooling flow and of the
fate of the cooled  material is of great importance for understanding
the evolution of the core of a cluster.

Edge et al. (1992) stress that the characterization of a cluster as a 
cooling flow depends heavily on the spatial 
resolution and sensitivity of the instruments used to image the cluster.
This was the main limitation for early work.

{\it ROSAT} offers, so far,  the best opportunity to resolve the 
cores of the clusters and thus establish in a reliable manner the
fraction of cooling flows in a flux-limited sample. 
It also offers a unique opportunity to study the 
occurrence of substructure in cooling flows (Peres, Buote \& Fabian, in preparation)
and to resolve mass deposition and cooling time profiles. 
The low background contamination of one of its
detectors, the Position Sensitive 
Proportional Counter (PSPC) allows surface
brightness profiles to be extracted to a large radius.
The good image capabilities of the other instrument, the 
High Resolution Imager (HRI), 
is important for the study of substructure and, in the context of
the present work, fundamental for determining the properties of 
more distant flows. 
  
In the absence of a volume-limited sample,  
statistical studies of cooling flows are best pursued with 
the use of a flux-limited sample, such as the one presented by Edge et
al. (1990).
The clusters in this sample are the brightest 55 over the sky in the 2-10 keV band 
(B55 sample henceforth), and all have fluxes 
above 1.7 $\times$ 10$^{-11}$ \ergpcmsqps. The clusters were selected from 
observations with the Einstein and EXOSAT observatories 
and the HEAO-1 and Ariel-V satellites.   

The work of Ebeling et al. (1996) on the X-ray Brightest Abell Cluster Sample
(XBACS) showed that the completeness the B55 sample was satisfactory.
A remaining issue in the analysis of cooling flow properties in this sample
was the ability to image the cooling region in the centres of clusters.
Here we analyse {\it ROSAT} pointed observations  of the clusters in the
B55 sample and deproject their surface brightness profiles. 
This allows us to build a catalogue of the ICM and cooling flow properties
of the B55 clusters, and to compare these to recent optical and radio 
data.
The structure of the paper is as follows. 
In sections 2 and 3 we describe the selection of the observations 
and the method of analysis, respectively. 
In section 4 we present our results, and in section 5 we summarize our conclusions. 
We assume throughout that
{\it H}$_0$ = 50 km s$^{-1}$ Mpc$^{-1}$ and  {\it q}$_0$ = 0.5.

\section{Observations}
\label{obs}

We have used archival and proprietary observations of the B55 clusters carried out 
with the Position Sensitive Proportional Counter (PSPC) 
and the High Resolution Imager (HRI) on {\it ROSAT}. 
(The only exceptions were the cases of 
3C129 and A2147 for which no useful ROSAT observations exist 
).
The PSPC provides a FWHM spatial resolution of $\sim$ 25 arcsec
(corresponding to a scale of 60 kpc for objects at a redshift
$z \sim$ 0.1) and a FWHM spectral resolution of 
$\frac{\Delta E}{E} = 0.43 \left(\frac{E}{0.93 keV} \right)^{-0.5}$,
whereas the HRI has a FWHM spatial resolution of $\sim$ 4 arcsec and
basically no spectral resolution.
 
The HRI is ideal for characterizing mass deposition profiles 
due to its high spatial resolution, whereas the PSPC is invaluable 
to recover ICM properties from the cluster surface brightness, 
due to its low and well characterized  background. 
In this work we have analysed 47 ROSAT PSPC observations, 
38 ROSAT HRI observations (32 observations with both 
instruments), covering a total of 85 images of 53 clusters.
When more than one observation existed for the 
same object with the same instrument, we used only   
the observation with the longest exposure time.
(Only in a few cases were  other criteria, e.g. the existence of a better 
centred image, considered more important.) 

A list of the clusters in the B55 sample, together with relevant 
generic information, is displayed in Tables \ref{info} and \ref{input}.  
The X-ray positions given in Table \ref{info} correspond to the peak 
in X-ray emission from our data, and were found by visual inspection. 
The optical positions were determined from  optical counterparts  
in images retrieved from the 
Space Telescope Digitized Sky Survey (DSS).
(We have used the HRI images, whenever available, to determine the 
locus of the X-ray peak.)
For cases where no clear single X-ray peak existed,
like A1367 and A1736, the centroid of the X-ray emission was used. 
The close agreement between the optical and X-ray positions 
can be seen from their offset, $\Delta \theta$, presented in Table 
\ref{info} (cf. sec. \ref{optical/radio}).

The radio information in Table \ref{info} was obtained by
searching the Green Bank 1400 and 4850MHz surveys 
(available through NED) and the 1400MHz FIRST and NVSS online catalogues.  
The information on the existence of optical emission lines was 
obtained from a variety of sources in the literature and from 
private communication (C.S. Reynolds and C.S. Crawford).

The X-ray observations used are identified by their 
Rosat Observation Request Sequence Number (ROR) in Table 
\ref{input}, where the raw exposure times and cluster redshifts for each 
observation are also listed. 
The other entries in this table are discussed in the next section.

%
\begin{table*}
\caption{{\bf B55 sample}.
(a) Name of the cluster; 
(b) Right ascension of X-ray peak (J2000);
(c) Declination of X-ray peak (J2000); 
(d) Right ascension of optical counterpart (J2000);
(e) Declination of optical counterpart (J2000);
(f) Offset between the X-ray and optical peaks, in arcsec; 
(g) Signal the presence ($\surd$), absence ($\times$), 
or lack of information ($-$) of optical emission lines;
(h) Radio flux density (in mJy) at 6 cm from the Green Bank survey;
(i) Radio flux density (in mJy) at 20 cm from the 
Green Bank ($\dagger$), NVSS or FIRST (*) surveys.
The values presented in columns (h) and (i) are for sources which are coincident with 
the optical or X-ray position within the angular resolution of the survey.
We follow the convention used in (g) to indicate lack of information on radio fluxes. }
\label{info}
\begin{center}
\begin{tabular}{crrcrrcccccc}
Cluster & R.A. (X-ray) & Dec. (X-ray) & R.A. (opt.) & Dec. (opt.) & $\Delta \theta$ &
Opt. Lines & Flux (5~GHz) & Flux (1.4~GHz) \\ 
\hline
(a) & (b) & (c) & (d) & (e) & (f) & (g) & (h) & (i) \\ 
\\A85 & 00 41 50.8 & $-$09 18 07 & 00 41 50.4 & $-$09 18 12 & 8 & $\surd$ & (4.6$\pm$1.1)$\times$10 & (5.8$\pm$0.3)$\times$10 
\\A119 & 00 56 16.8 & $-$01 14 45 & 00 56 16.1 & $-$01 15 19 & 36 & $\times$ & $-$ & $\times$ 
\\A262 & 01 52 45.4 & 36 09 26 & 01 52 46.5 & 36 09 06 & 24 & $\surd$ & $-$ & (1.31)$\times$10$^2$ $^{\dagger}$
\\AWM7 & 02 54 27.4 & 41 34 51 & 02 54 27.5 & 41 34 46 & 5 & $\times$ & $-$ & $-$ 
\\A399 & 02 57 53.6 & 13 01 47 & 02 57 53.2 & 13 01 50 & 7 & $\times$ & $-$ & $\times$
\\A401 & 02 58 56.0 & 13 35 03 & 02 58 57.8 & 13 34 57 & 27 & $\times$ & $-$ & $\times$
\\A3112 & 03 17 57.7 & $-$44 14 17 & 03 17 57.7 & $-$44 14 18 & 1 & $\times$ & (6.9$\pm$0.4)$\times$10$^2$ & $-$
\\A426 & 03 19 48.0 & 41 30 46 & 03 19 48.3 & 41 30 41 & 6 & $\surd$ & (4.2$\pm$0.5)$\times$10$^4$ & (2.12)$\times$10$^4$ $^{\dagger}$ 
\\2A 0335+096 & 03 38 40.2 & 09 58 12 & 03 38 40.6 & 09 58 11 & 6 & $\surd$ & $-$ & (2.41$\pm$0.12)$\times$10 
\\A3158 & 03 42 50.9 & $-$53 37 32 & 03 42 53.0  & $-$53 37 53 & 28 & $\times$ & $-$ & $-$ 
\\A478 & 04 13 25.0 & 10 27 59 & 04 13 25.3 & 10 27 54 & 7 & $\surd$ & $-$ & (3.55$\pm$0.15)$\times$10 
\\A3266 & 04 31 15.7 & $-$61 27 08 & 04 31 13.5  & $-$61 27 12 & 11 & $\times$ & $-$ & $-$
\\A496 & 04 33 37.6 & $-$13 15 40  & 04 33 37.8 & $-$13 15 43 & 4 & $\surd$ & (4.4$\pm$1.1)$\times$10 & $-$ 
\\3C129 & ------------  & -----------  & 04 48 58.2 & 45 02 01  & $-$ & $-$ & $-$ & (5.31)$\times$10$^3$ &    
\\A3391 & 06 26 19.9 & $-$53 41 53 & 06 26 20.4 & $-$53 41 36 & 18 & $\times$ & (1.922$\pm$0.019)$\times$10$^3$ & $-$ 
\\A576 & 07 21 31.2  & 55 45 52 & 07 21 30.2 & 55 45 40 & 15 & $\times$ & $-$ & $\times$ 
\\PKS 0745-191 & 07 47 30.9 & $-$19 17 43 & 07 47 31.3 & $-$19 17 40 & 6 & $\surd$ & (4.8$\pm$0.3)$\times$10$^2$ & (2.37$\pm$0.08)$\times$10$^3$
\\A644 & 08 17 25.5 & $-$07 30 40 & 08 17 25.6 & $-$07 30 46 & 6 & $\times$ & $-$ & $-$ 
\\A754 & 09 09 18.8 & $-$09 41 20 & 09 08 32.4 & $-$09 37 49 & 691 & $\times$ & $-$ & (9.0$\pm$1.0)$\times$10$^0$ 
\\Hyd-A & 09 18 05.8  & $-$12 05 40 & 09 18 05.6 & $-$12 05 44 & 5 & $\surd$ & (1.40$\pm$0.01)$\times$10$^4$ & (4.08$\pm$0.13)$\times$10$^4$ 
\\A1060 & 10 36 43.2 & $-$27 31 40 & 10 36 42.8 & $-$27 31 41 & 6 & $\surd$ & $-$ & $\times$ 
\\A1367 & 11 44 48.2  & 19 42 05 & 11 44 48.0 & 19 41 18 & 47 & $\times$ & $-$ & $\times$ 
\\Virgo & 12 30 49.0 & 12 23 35 & 12 30 49.4 & 12 23 26 & 9 & $\surd$ & (6.1$\pm$0.8)$\times$10$^4$ & (2.24)$\times$10$^4$ $^{\dagger}$ 
\\Cent & 12 48 48.9 & $-$41 18 44 & 12 48 49.1 & $-$41 18 42 & 3 & $\surd$ & (1.53$\pm$0.08)$\times$10$^3$ & $-$ 
\\Coma & 12 59 35.6 & 27 57 31  & 12 59 35.6 & 27 57 34 & 6 & $\times$ & (8.4$\pm$1.2)$\times$10 & (2.07$\pm$0.07)$\times$10$^2$ 
\\A1644 & 12 57 12.2 & $-$17 24 34 & 12 57 11.6 & $-$17 24 35 & 9 & $\times$ & (1.12$\pm$0.12)$\times$10$^2$ & (9.9$\pm$0.3)$\times$10
\\A3532 & 12 57 21.8 & $-$30 21 51 & 12 57 22.0 & $-$30 21 50 & 3 & $-$ & (4.4$\pm$0.3)$\times$10$^2$ & (1.16$\pm$0.04)$\times$10$^3$
\\A1650 & 12 58 41.7 & $-$01 45 44 & 12 58 41.5 & $-$01 45 41 & 4 & $-$ & $-$ & $-$  
\\A1651 & 12 59 21.7 & $-$04 11 47 & 12 59 22.5 & $-$04 11 46 & 12 & $\times$ & $-$ & $-$ 
\\A1689 & 13 11 29.5 & $-$01 20 28 & 13 11 29.5 & $-$01 20 29 & 1 & $\times$ & $-$ & $\times$ 
\\A1736 & 13 26 50.0  & $-$27 10 20  & 13 26 48.7 & $-$27 08 37 & 103 & $\times$ & $-$ & $\times$ 
\\A3558 & 13 27 56.5 & $-$31 29 44 & 13 27 56.8 & $-$31 29 45 & 4 & $\times$ & $-$ & (4.5$\pm$0.5)$\times$10$^0$ 
\\A3562 & 13 33 36.0 & $-$31 40 05  & 13 33 34.7 & $-$31 40 21 & 23 & $\times$ & $-$ & $\times$ 
\\A3571 & 13 47 28.4 & $-$32 51 55 & 13 47 28.3 & $-$32 51 55 & 1 & $-$ & $-$ & (8.4$\pm$1.7) 
\\A1795 & 13 48 52.7 & 26 35 30 & 13 48 52.6 & 26 35 35 & 5 & $\surd$ & (2.6$\pm$0.3)$\times$10$^2$ & (9.3$\pm$0.3)$\times$10$^2$ $^*$ 
\\A2029 & 15 10 55.8 & 05 44 46 & 15 10 56.1 & 05 44 41 & 7 & $\times$ & (8.9$\pm$1.4)$\times$10 & (5.5)$\times$10$^2$ $^{\dagger}$ 
\\A2052 & 15 16 43.7 & 07 01 19 & 15 16 44.6 & 07 01 17 & 13 & $\surd$ & (1.03$\pm$0.14)$\times$10$^3$ & (5.4)$\times$10$^3$ $^{\dagger}$ 
\\MKW3s & 15 21 51.8 & 07 42 24 & 15 21 51.9 & 07 42 30 & 6 & $\surd$ & $-$ & (1.26)$\times$10$^2$ $^{\dagger}$ 
\\A2065 & 15 22 29.0 & 27 42 33 & 15 22 29.2 & 27 42 26 & 7 & $\times$ & $-$ & (1.4$\pm$0.3)$\times$10  
\\A2063 & 15 23 04.8 & 08 36 20 & 15 23 05.3 & 08 36 33 & 15 & $\times$ & $-$ & (1.5$\pm$0.07)$\times$10 
\\A2142 & 15 58 20.2 & 27 13 52 & 15 58 20.1 & 27 14 00 & 8 & $\surd$ & $-$ & $\times$
\\A2147 & -----------  & ----------- & 16 02 17.0 & 15 58 27 & $-$ & $\times$ & $-$ & $-$  
\\A2199 & 16 28 37.7 & 39 33 03  & 16 28 38.6 & 39 33 04 & 7 & $\surd$ & (4.8$\pm$0.5)$\times$10$^2$ & (3.7)$\times$10$^3$ $^{\dagger}$ 
\\A2204 & 16 32 47.1 & 05 34 34 & 16 32 46.8 & 05 34 31 & 5 & $\surd$ & $-$ & (7.01$\pm$0.2)$\times$10 
\\Tri Aust & 16 38 20.3 & $-$64 21 28 & 16 38 18.3 & $-$64 21 36 & 20 & $-$ & $-$ & $-$ 
\\A2244 & 17 02 41.9 & 34 03 30 & 17 02 42.5 & 34 03 35 & 9 & $\times$ & $-$ & (2.41$\pm$0.13)$\times$10$^0$ $^*$ 
\\A2256 & 17 03 13.9 & 78 39 06 & 17 04 27.1 & 78 38 25 & 59 & $\times$ & $-$ & $\times$ 
\\Ophiuchus & 17 12 27.8 & $-$23 22 08 & 17 12 28.2 & $-$23 22 09 & 6 & $\surd$ & $-$ & (2.91$\pm$0.1)$\times$10
\\A2255 & 17 12 36.2 & 64 04 09 & 17 12 35.0 & 64 04 14 & 9 & $\times$ & $-$ & $\times$
\\A2319 & 19 21 09.7 & 43 56 48 & 19 21 10.1 & 43 56 43 & 7 & $\times$ & $-$ & $\times$ 
\\Cyg-A & 19 59 28.1 & 40 44 05 & 19 59 28.4 & 40 44 01 & 5 & $\surd$ & (2.1$\pm$0.3)$\times$10$^5$ & $-$
\\A3667 & 20 12 24.3 & $-$56 49 49 & 20 12 27.4 & $-$56 49 37 & 25 & $\times$ & $-$ & $-$
\\A2597 & 23 25 19.3 & $-$12 07 20 & 23 25 19.7 & $-$12 07 27 & 9 & $\surd$ & (4.1$\pm$0.2)$\times$10$^2$ & (1.88$\pm$0.06)$\times$10$^3$ 
\\Klem44 & 23 47 43.4 & $-$28 08 20 & 23 47 43.4 & $-$28 08 37 & 17 & $-$ & $-$ & (2.84$\pm$0.13)$\times$10
\\A4059 & 23 57 00.2 & $-$34 45 39 & 23 57 00.5 & $-$34 45 35 & 5 & $\surd$ & (1.17$\pm$0.13)$\times$10$^2$ & (1.29$\pm$0.04)$\times$10$^3$
\end{tabular}
\end{center}
\end{table*}           
%
%
%
%
\begin{table*}
\caption{{\bf Input parametres for the deprojection analysis}.
(a) Name of the cluster. The letter inside brackets indicate wether 
the observation was made with the PSPC(P), or with the HRI(H); 
(b) Rosat Observation Request Sequence Number (ROR); 
(c) Raw exposure time in seconds; 
(d) Redshift of the cluster; 
(e) Bin size (in kpc) used in the deprojection;
(f) Galactic neutral hydrogen column density along the line of sight to 
the cluster, given in units of 10$^{21}$ cm$^{-2}$; 
(g) Temperature (in keV) from the catalogues of David et al. (1993) and White et al. (1997);
(h) Outer radius to which the surface brightness profile was 
extracted (in Mpc).
}
\label{input}
\begin{center}
\begin{tabular}{clrrrrrcccccc}
Cluster & ROR & $\Delta t$ & $z$ & bin size & $N_{\rm H}$ & 
$\langle {\rm k}T \rangle$ & $R_{\rm out}$ \\
\hline
(a) & (b) & (c) & (d) & (e) & (f) & (g) & (h) \\ 
\\ A85(H) & rh800271 & 17308 & 0.0521 & 11.1 & 0.3 & 6.2 & 0.244 
\\ A85(P) & rp800250 & 10240 & 0.0521 & 41.5 & 0.3 & 6.2 & 0.914 
\\ A119(P) & rp800251 & 15203 & 0.044 & 35.6 & 0.35 & 5.1 & 1.049 
\\ A262(P) & rp800254 & 8719 & 0.0164 & 20.9 & 0.53 & 2.4 & 0.452 
\\ AWM7(H) & wh800364 & 14864 & 0.0172 & 3.8 & 0.92 & 3.6 & 0.140 
\\ AWM7(P) & wp800168 & 13335 & 0.0172 & 14.6 & 0.92 & 3.6 & 0.532 
\\ A399(H) & rh800850n00 & 6833 & 0.0715 & 73.5 & 1.17 & 5.8 & 0.757 
\\ A401(P) & rp800235 & 7465 & 0.0748 & 57.3 & 1.11 & 7.8 & 1.175 
\\ A3112(P) & rp800302n00 & 7600 & 0.0746 & 57.2 & 0.4 & 4.1 & 0.886
\\ A3112(H) & rh800627a01 & 9540 & 0.0746 & 16.0 & 0.4 & 4.1 & 0.421
\\ A426(H) & wh800068 & 10785 & 0.0183 & 5.93 & 1.45 & 5.5 & 0.090 
\\ A426(P) & wp800186 & 4787 & 0.0183 & 23.2 & 1.45 & 5.5 & 0.766 
\\ 2A 0335+096(H) & rh800050 & 14012 & 0.0349 & 5.14 & 1.72 & 3.0 & 0.190 
\\ 2A 0335+096(P) & wp800083 & 10346 & 0.0349 & 28.7 & 1.72 & 3.0 & 0.731 
\\ A3158(P) & rp800310 & 3022 & 0.0575 & 68.1 & 0.12 & 5.5 & 1.022 
\\ A478(H) & wh800091 & 22712 & 0.0882 & 17.6 & 1.36 & 6.8 & 0.520 
\\ A478(P) & wp800193 & 22139 & 0.0882 & 66.0 & 1.36 & 6.8 & 1.420 
\\ A3266(P) & wp800552n00 & 13560 & 0.0594 & 70.1 & 0.3 & 6.2 & 0.678 
\\ A3266(H) & rh800628n00 & 8202 & 0.0594 & 25.0 & 0.3 & 6.2 & 0.608
\\ A496(H) & rh800272 & 14493 & 0.033 & 7.2 & 0.44 & 4.7 & 0.236 
\\ A496(P) & rp800024 & 8972 & 0.033 & 27.2 & 0.44 & 4.7 & 0.476 
\\ A3391(P) & wp800080 & 6781 & 0.0545 & 64.9 & 0.45 & 5.2 & 0.930 
\\ A576(H) & rh800727n00 & 10090 & 0.0381 & 16.6 & 0.56 & 4.3 & 0.236 
\\ PKS 0745-191(H) & wh800398n00 & 23385 & 0.1028 & 20.0 & 4.66 & 8.5 & 0.330 
\\ PKS 0745-191(P) & wp800623n00 & 10477 & 0.1028 & 75.1 & 4.66 & 8.5 & 1.464 
\\ A644(H) & rh800273n00 & 18668 & 0.0704 & 21.7 & 0.73 & 6.6 & 0.355 
\\ A644(P) & wp800379n00 & 10285 & 0.0704 & 81.5 & 0.73 & 6.6 & 0.897 
\\ A754(P) & rp800232n00 & 6359 & 0.0542 & 64.6 & 0.47 & 8.7 & 1.270 
\\ A754(H) & rh800768a01 & 37346 & 0.0542 & 23.0 & 0.47 & 8.7 & 0.466
\\ HYD-A(H) & rh800132 & 27488 & 0.0522 & 5.21 & 0.48 & 3.8 & 0.100 
\\ HYD-A(P) & rp800318n00 & 18403 & 0.0522 & 62.4 & 0.48 & 3.8 & 0.894 
%
%
%
%
%
%
%
%
\\ A1060(P) & wp800200 & 15852 & 0.0124 & 21.2 & 0.5 & 3.3 & 0.365 
\\ A1060(H) & rh800632n00 & 14967 & 0.0124 & 11.0 & 0.5 & 3.3 & 0.160 
\\ A1367(P) & rp800153 & 18982 & 0.0215 & 45.2 & 0.22 & 3.5 & 0.533 
\\ VIRGO(H) & wh700214 & 14239 & 0.0037 & 1.2 & 0.25 & 2.4 & 0.050 
\\ VIRGO(P) & wp800187 & 10539 & 0.0037 & 4.8 & 0.25 & 2.4 & 0.159 
\\ CENT(H) & rh700320a01 & 16526 & 0.0109 & 4.8 & 0.8 & 3.6 & 0.090 
\\ CENT(P) & wp800192 & 7985 & 0.0109 & 14.0 & 0.8 & 3.6 & 0.331 
\\ COMA(P) & rp800005 & 22183 & 0.0232 & 29.2 & 0.09 & 8.0 & 0.865 
\\ COMA(H) & rh800242a04 & 37410 & 0.0232 & 16.0 & 0.09 & 8.0 & 0.250 
\\ A1644(H) & rh800851a01 & 10232 & 0.0474 & 15.2 & 0.47 & 4.7 & 0.137 
\\ A3532(P) & wp701155n00 & 8620 & 0.0585 & 69.2 & 0.62 & 4.4 & 0.876 
\\ A1650(H) & rh800852a01 & 6054 & 0.0845 & 25.5 & 0.15 & 5.5 & 0.297 
\\ A1651(P) & wp800353 & 7435 & 0.0846 & 63.7 & 0.17 & 7.0 & 1.243 
\\ A1689(P) & rp800248 & 13957 & 0.181 & 115.6 & 0.19 & 10.1 & 0.982 
\\ A1689(H) & rh800445n00 & 13094 & 0.181 & 33.0 & 0.19 & 10.1 & 0.823
\\ A1736(H) & rh800853a01 & 13783 & 0.046 & 29.6 & 0.5 & 4.6 & 0.291 
\\ A3558(H) & wh800399 & 16792 & 0.0478 & 25.6 & 0.45 & 6.5 & 0.409 
\\ A3558(P) & wp800076 & 30213 & 0.0478 & 96.0 & 0.45 & 6.5 & 1.593 
\\ A3562(P) & rp800237n00 & 20202 & 0.0499 & 59.9 & 0.42 & 3.8 & 1.297 
\\ A3571(H) & rh800626n00 & 19460 & 0.0391 & 25.5 & 0.4 & 7.6 & 0.446 
\\ A3571(P) & rp800287 & 6072 & 0.0391 & 31.9 & 0.40 & 7.6 & 0.781 
\\ A1795(P) & rp800105n00 & 36273 & 0.0627 & 49.1 & 0.12 & 5.1 & 0.957 
\\ A1795(H) & rh800222a01 & 11097 & 0.0627 & 13.0 & 0.12 & 5.1 & 0.506
\end{tabular}
\end{center}
\end{table*}
%
%
\begin{table*}{{\bf Table \ref{input} - }continued.}
\begin{center}
\begin{tabular}{clrrrrrcccccc}
Cluster & ROR & $\Delta t$ & $z$ & bin size & $N_{\rm H}$ & 
$\langle {\rm k}T \rangle$ & $R_{\rm out}$ \\
\hline
(a) & (b) & (c) & (d) & (e) & (f) & (g) & (h) \\
\\ A2029(H) & rh150024 & 17757 & 0.0767 & 15.6 & 0.31 & 7.8 & 0.383 
\\ A2029(P) & rp800249 & 12550 & 0.0767 & 58.6 & 0.31 & 7.8 & 1.025 
\\ A2052(P) & rp800275 & 6215 & 0.0348 & 57.2 & 0.29 & 3.4 & 0.557 
\\ A2052(H) & rh800223n00 & 4429 & 0.0348 & 15.0 & 0.29 & 3.4 & 0.308 
\\ MKW3(P) & rp800128 & 9996 & 0.0449 & 72.5 & 0.29 & 3.0 & 0.562 
\\ MKW3(H) & rh800425n00 & 13502 & 0.0449 & 10.0 & 0.29 & 3.0 & 0.244 
\\ A2065(H) & rh800724n00 & 11780 & 0.0722 & 22.2 & 0.29 & 8.4 & 0.363 
\\ A2063(P) & wp800184 & 10198 & 0.0350 & 22.2 & 0.29 & 4.1 & 0.363 
\\ A2142(P) & wp150084 & 7740 & 0.0899 & 67.1 & 0.39 & 11 & 1.443 
\\ A2142(H) & rh800640n00 & 19785 & 0.0899 & 27 & 0.39 & 11 & 0.910
\\ A2199(H) & wh800071 & 5222 & 0.030 & 19.9 & 0.09 & 4.7 & 0.195 
\\ A2199(P) & wp800644n00 & 41082 & 0.030 & 24.8 & 0.09 & 4.7 & 0.733 
\\ A2204(H) & rp800750n00 & 15488 & 0.1523 & 40.9 & 0.56 & 9.0 & 0.667
\\ A2204(P) & rp800281 & 5359 & 0.1523 & 102.2 & 0.56 & 9.0 & 1.175 
\\ TRI AUST(P) & rp800280n00 & 7338 & 0.051 & 40.7 & 1.98 & 7.9 & 0.916 
\\ A2244(P) & rp800265n00 & 2965 & 0.1024 & 112.2 & 0.2 & 7.1 & 0.935
\\ A2256(P) & wp100110 & 17865 & 0.0581 & 45.8 & 0.43 & 7.5 & 1.169 
\\ A2256(H) & rh800676n00 & 43567 & 0.0581 & 25.0 & 0.43 & 7.5 & 0.745
\\ OPHI(P) & rp800279n00 & 3932 & 0.028 & 23.3 & 1.97 & 9.0 & 0.756 
\\ OPHI(H) & wh800067 & 22285 & 0.028 & 9.3 & 1.97 & 9.0 & 0.202 
\\ A2255(P) & rp800512n00 & 14555 & 0.0809 & 92.0 & 0.26 & 7.3 & 1.258 
\\ A2319(H) & wh800072 & 5559 & 0.0564 & 29.7 & 0.86 & 9.9 & 0.303 
\\ A2319(P) & wp800073a01 & 3171 & 0.0564 & 44.6 & 0.86 & 9.9 & 1.138 
\\ CYG-A(H) & rh800021 & 43240 & 0.057 & 24.0 & 3.61 & 7.3 & 0.390 
\\ CYG-A(P) & wp800622n00 & 9447 & 0.057 & 67.6 & 3.61 & 7.3 & 1.464 
\\ A3667(P) & rp800234n00 & 12560 & 0.0530 & 42.2 & 0.4 & 6.5 & 1.328 
\\ A2597(H) & rh800111 & 17996 & 0.0824 & 24.9 & 0.25 & 6.0 & 0.241 
\\ A2597(P) & rp800112 & 7243 & 0.0824 & 62.3 & 0.25 & 6.0 & 0.904 
\\ KLEM44(P) & wp800354n00 & 3353 & 0.0283 & 23.5 & 0.15 & 3.3 & 0.505 
\\ A4059(H) & rh800224n00 & 6320 & 0.0478 & 15.3 & 0.11 & 3.5 & 0.190 
\\ A4059(P) & wp800175 & 5514 & 0.0478 & 38.4 & 0.11 & 3.5 & 0.595 
\end{tabular}
\end{center}
\end{table*}

\section{METHOD OF ANALYSIS}
\label{method}

We have undertaken the present analysis in two distinct stages: (i) the
reduction of the data to produce one image per observation, and
(ii) the deprojection of the azimuthally averaged surface brightness
profiles extracted from these images.

In the first stage, PSPC images have been constructed from counts in
the energy range 0.4-2.0 keV and have been vignetting
corrected. They have also been corrected for instrument and telemetry dead time
and had periods of high particle background and scattered solar X-rays
removed. With the selection of this range of energies we minimize the
Galactic  and particle backgrounds (which dominate below 0.4 keV and
above 2.1 keV, respectively), while retaining most of the cluster emission. 
The vignetting and exposure-time corrections were performed 
using the IDL routines {\it make\_emap.pro}, 
{\it make\_image.pro}, {\it mexdiv.pro}; this follows the prescription
of Snowden et al. (1994). The background subtraction of PSPC images was 
left to the deprojection software, written by D. White and collaborators. 
(We select the bin for which the background counts are accumulated 
by inspection of the azimuthally averaged surface brightness, which works very well 
for the PSPC due to its low background count rate.)


The HRI images were reduced in a different manner. Background subraction was 
performed with the STARLINK software ASTERIX, by accumulating counts in 
circles of approximately 0.05 degrees in a clear off-centre area of the image.
The vignetting correction was left to the deprojection software. 
In all  cases we were careful to inspect each background-subtracted  
surface-brightness profile and only deproject it up to a point where 
the difference between the subtracted and non-subtracted values  differ insignificantly. 
In this way we guarantee that background subtraction does not affect 
the results significantly.   

For both the PSPC and HRI images we have identified
contaminating sources by eye and masked them out 
(we have also masked out the detector supporting structure in the PSPC images). 
Finally, we have  extracted the profile of counts in annuli 
around the peak of the X-ray emission, to obtain 
a surface brightness profile suitable for deprojection. 
All the image analysis was carried out with ASTERIX and our deprojection software
%
%
%

In the second stage we use the single-phase algorithm for surface
brightness deprojection of
Fabian et al. (1980), first used for the analysis of clusters of
galaxies by Fabian et al. (1981)\footnote{A more sophisticated
deprojection, taking into account the multiphase nature of the ICM is
known to give results in agreement with our simple approach here
(Thomas et al. 1987).}. 
In this technique counts in an X-ray 
image are assumed to  come from a spherically symmetric 
(single phase) hot thin plasma in  hydrostatic equilibrium. 
Modelling the cluster emission as such, allows us to
obtain the emission from each spherical shell in the cluster from the
observed surface brightness profile and to compare this emission with the
predicted value from a plasma code. (For this work we have used the
MEKAL code; cf. Mewe et al 1985, 1986, Kaastra 1992, and Arnaud et al. 1985.) 
Through this comparison we can obtain the radial distribution 
of the thermodynamic properties of
the gas in the ICM, 
namely temperature, pressure and electron
density. Once these quantities are known, secondary properties like
luminosity, central cooling time and mass deposition rates can be
estimated. A more detailed exposition of the deprojection
algorithm used here can be found in White, Jones \& Forman (1997)
and references therein.

The input parameters in the implementation
of the deprojection algorithm used here are: redshift of the cluster,
average temperature (as measured by Einstein MPC, EXOSAT or GINGA; David et al. 1992), 
spectral band of the instrument (0.1-2.4 keV), Galactic 
hydrogen column density along the cluster line of sight, 
binsize of the data, {\it H}$_0$  and {\it q}$_0$, pressure at the outermost bin,
cluster core radius and velocity dispersion, 
galaxy linear mass (GLM), and 
galaxy linear mass cut-off (GCTOF)
\footnote{The gravitational model for the 
galaxy has a linearly increasing  mass profile and a cut-off 
radius to avoid runaway. This is analogous to the model assumed by Thomas et al. (1987).}. 
From these, only four are kept free to fit the X-ray data: 
the cluster core radius  and velocity 
dispersion, GLM and GCTOF~\footnote{Two caveats must be mentioned here: (1) Although 
we keep the value of the velocity dispersion free, the values used are
roughly in agreement with (at least the lower limits of) 
the values measured by optical studies (cf. Zabludoff et al. 1990, Fadda et al. 1996).
(2) The pressure at the outermost bin is not considered as a free parameter since it is 
always adjusted for the temperature at the outermost bin to agree roughly with the
averaged value. 
(3) The parameters of the central cluster galaxy
do not sensitively affect the results presented here in most cases, 
due to the resolution of the data.}. 
We model the potential as an isothermal sphere, adjusting
the cluster core radius and velocity dispersion to conform to this model.
This follows the works of Allen et al. (1996) and Fukazawa et al. (1994).
The latter showed that the hot ICM phase exists up to the central
parts of clusters of galaxies while the former showed that 
a consistent modeling of the cluster potential is achieved by assuming
a roughly constant average temperature. This is discussed at 
length in Sec. \ref{multiphase}.

Galactic absorption was accounted for by using 
the hydrogen column densities quoted by Stark et al. (1992).
No excess column density was introduced in our analysis, 
although this was found to exist from the X-ray spectroscopic studies  
of cooling flow clusters (White et al. 1991, Allen \& Fabian 1997).
Any excess absorption increases the mass deposition rates 
since the  
luminosity must be increased to produce the observed counts.

We have usually rebinned the PSPC and HRI data
to 30 and 8 arcsec, respectively to minimize  
complexities associated with the instrument PSF (there are cases with 
coarser binning. The exact bin size used for each case can be found in 
Table \ref{input}).
Most of the values used for the average temperature 
were taken from  David et al. (1992), although more recent 
data were used when available.
The value used for A2597 is  6.0 keV.
 
We have used a Monte-Carlo algorithm to perturb and deproject 
each surface brightness profile 100 times. By doing so we can 
place limits on the quantities derived from the deprojection analysis.
However, we stress that these limits should be associated with the 
assumed potential only. Systematic errors in the potential 
will lead to larger uncertainties in the derived quantities.

As a final remark we note that the ICM is assumed to
be in hydrostatic equilibrium. This is a key assumption needed to integrate
the cooling flow equations, once a gravitational potential is
assumed. It is reasonable since the sound crossing time is usually 
shorter than the age of the cluster, which we assume to be 
$\sim$ {\it H}$_0^{-1}$.
Supportive evidence for this and the hypothesis that the temperature profile
is roughly constant is given below.

\subsection{The Multiphase (Inhomogeneous) Intracluster Medium}
\label{multiphase}

Suggestive evidence that the highest temperature does not vary much with radius 
appeared in the work of Fukazawa et al. (1994), where the Centaurus cluster 
was observed with the ASCA satellite. They showed that  hot components 
of the ICM exist throughout the central 5 arcmin ($\sim$~130 kpc) of this cluster.
By fitting a two temperature Raymond-Smith(RS) model to the data 
these authors showed that the temperature profile for the
cluster is approximately constant up to the 
inner $\sim$~2 arcmin ($\sim$50 kpc). 
A study of the cluster of galaxies A1060 by Tamura et al. (1996)
reached similar conclusions.

The hypothesis that the cluster temperature profile can be modeled  as approximately
flat received definitive support from the work of Allen et al. (1996). 
The authors studied the cluster PKS 0745-191 (0745 henceforth)
with X-ray (spectral and imaging) and gravitational lensing methods. 
By carrying out a spatial analysis similar to ours the authors showed
that the deprojected gravitational mass is in agreement 
with the mass inferred from the modelling of the lens provided the
temperature profile is assumed to be flat. 
In the study by Allen et al. (1996) additional supporting evidence for 
the assumption of a flat temperature profile was obtained by spectroscopy.
The X-ray spectrum of 0745, taken with  ASCA, 
was only reasonably fitted if a two-temperature or a CFLOW model 
\footnote{This is the same model used in this paper and described in 
Sec. 3. It is assumed that at each radial bin there is 
(i) gas in a single phase in hydrostatic equilibrium and 
(ii) gas cooling out of the flow.} 
were used.
From this analysis Allen et al. (1996) showed 
that once the effects of the cooling
flow are taken into account in the spectral analysis (through the CFLOW
model) the cluster is observed to be approximately isothermal. 
Comparing the X-ray and lensing results for 0745 the authors conclude that 
the temperature profile should remain approximately isothermal 
up to $\sim$10 arcsec (instead of the 2 arcmin limit obtained from the 
ASCA data alone).

From the work of Allen et al. (1996) it is clear that the assumption of an
isothermal profile for the gas makes the single-phase deprojection
analysis appropriate for modelling the true multi-phase nature of the ICM and
the cluster gravitational potential
\footnote{
Allen et al. (1996) did  not use a full multiphase deprojection method,  
resorting only to an additional cooling flow component (CFLOW)
in their spectral analysis. This is exactly the component 
present in our single phase deprojection algorithm, and it simulates 
a whole range of phases cooling out of the X-ray band.
Thomas et al. (1987) were able to  show, through a study of 11 clusters,
that the single-phase deprojection method gives results in
accordance with a multiphase treatment, where more than one phase
(apart from the material which is dropping out of the flow)
exists at each radial bin.
}
(gravitational lensing was used as a cross-check). 
For the clusters in this sample we cannot resort to the approaches discussed 
above to cross-check the appropriateness of our models, since
spatially resolved spectroscopic information is not available for 
most of the B55 clusters, and their redshifts and masses 
are not suitable for gravitational lensing studies.
However two indications of the appropriateness of our method are: 
(i) the size of the mass core radii we obtain for the B55 clusters and 
(ii) the functional form of the mass deposition rate profile inferred from the analysis.
   
\begin{figure}
\centerline{
\psfig{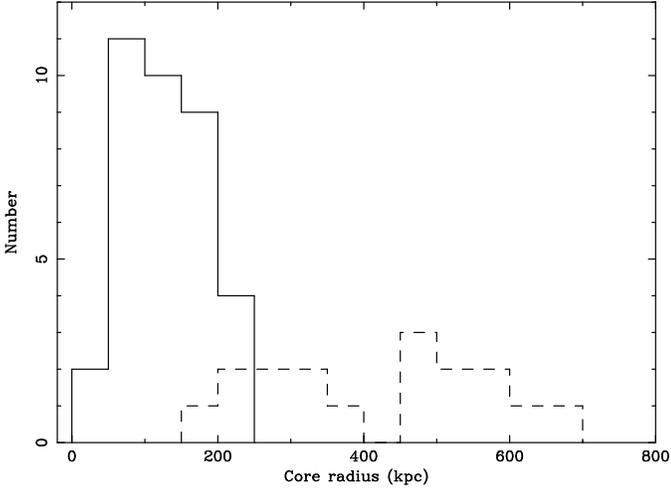}
}
\caption{
The distribution of core radii in the B55 sample shows the same bi-modality 
found by the X-ray and lensing studies of Allen (1997). 
Cooling flows have smaller core radii overall and show 
a distribution peaked around 100 kpc. Non-colling-flows display a broader
distribution of values, usually above 300 kpc. 
}
\label{core-rad}
\end{figure}

The values of mass core radii for the B55 clusters used in our analyses
are available from Table 3 and are plotted in Fig. \ref{core-rad}. 
They were obtained from our requirement that the mean X-ray gas temperature 
stay approximatelly constant as the evidence presented above suggests.
This methodology produces core radii which are 
overall smaller than ones found in previous analyses (White et al 1997, Edge et al. 1992)
where different assumptions were used.
Our findings agree with the trend of smaller core radii found from gravitational lensing 
studies of more distant clusters and with the 
multiphase picture of the ICM
(Fabian et al. 1984; see also recent work of Miralda-Escude \& Waxman 1995).

Our model of the cluster gravitational potential  includes a separate  mass component for 
the central galaxy. Despite this extra complication 
it is interesting to note from Fig. \ref{core-rad} that 
there is a clear separation between cooling-flow and non-cooling-flow clusters.
This is supported by the X-ray/lensing analysis of Allen (1997).


The widespread deposition of material over the cluster core also 
corroborates the multiphase nature of the ICM, where various different phases 
drop out of the flow at different radii. In fact simple 
multiphase models proposed by Nulsen (1987) find 
$\dot {M} (<{r}) \sim {r} ^{\alpha}$, with $\alpha \sim 1$
(cf. Sec. \ref{profiles}).


\begin{figure*}
\centerline{
\psfig{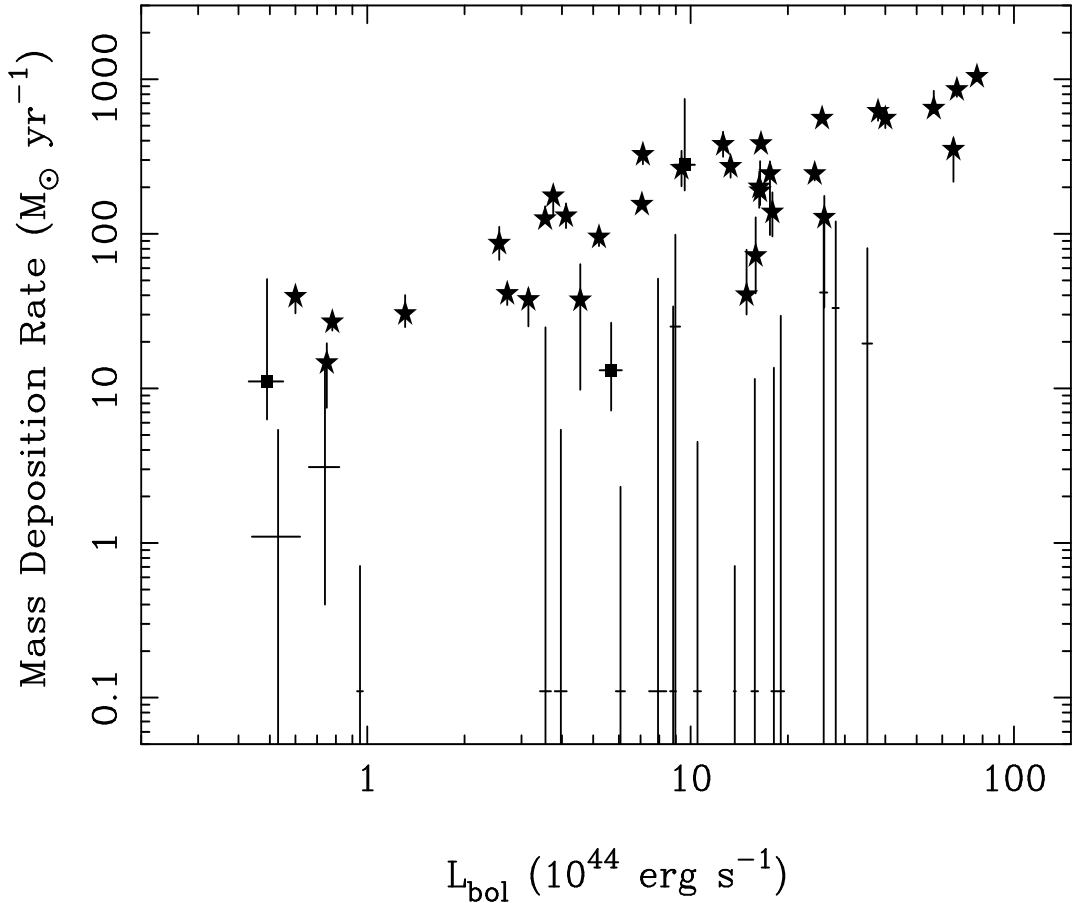}
}
\caption{
Mass deposition rate vs.
bolometric X-ray luminosity for the clusters in our sample. 
The lack of points in the upper left corner only reflects the maximum 
$\dot {M}$ allowed for a given luminosity.
Stars are used to denote PSPC observations. Squares represent the clusters for which only  
HRI observations were available. 
No special symbols are used to denote the clusters which have mass deposition rates
consistent with zero. The error bars for these clusters extend to the bottom of the figure 
and are kept to indicate the upper limit in their mass deposition 
rates (When the median mass deposition is zero we assume it equal to 0.1 for display
purposes).
The luminosity of the Virgo cluster was taken from White et al. (1997)
.}
\label{luminosity}
\end{figure*}


\section{RESULTS/DISCUSSION}
\label{sec:results}

The deprojected profiles of all clusters analyzed by us cannot 
be presented here due to space limitations. Instead we present
a list with the input parameters necessary to reproduce the analyses 
in Tables \ref{input}, and \ref{potential}. The results are 
displayed in a catalogue form in Table \ref{results} and discussed in
the next subsections.

\begin{table}
\caption{{\bf Parameters for modeling the cluster potential.}
(a) cluster name, 
(b) cluster velocity dispersion (in \kmps),
(c) cluster core radius (in kpc),
(d) linear galaxy mass (in 10$^{8}$ \Msunppc),
(e) cut-of for the galaxy mass (in Mpc). Note that the profile for the Coma 
cluster was fitted with a single isothermal model.
}
\label{potential}
\begin{center}
\begin{tabular}{crrcccc}
Cluster & $\sigma_{\rm deproj}$ & $a_{\rm deproj}$ & GLM & GCTOF\\
\hline
\\ A85 & 700 & 130 & 2.0 & 0.06 	          
\\ A119 & 680 & 600 & 1.0 & 0.08   	
\\ A262 & 450 & 130 & 1.2 & 0.04 	
\\ AWM7 & 520 & 200 & 1.1 & 0.08	
\\ A399 & 940 & 550 & 1.0 & 0.10 	
\\ A401 & 950 & 350 & 0.8 & 0.08  
\\ A3112 & 670 & 120 & 1.4 & 0.06   
\\ A426 & 700 & 60 & 0.9 & 0.06    
\\ 2A 0335+096 & 600 & 40 & 0.1 & 0.05  	
\\ A3158 & 860 & 400 & 0.8 & 0.08
\\ A478 & 850 & 180 & 1.8 & 0.08	
\\ A3266 & 780 & 500 & 1.5 & 0.17	
\\ A496 & 610 & 60 & 1.2 & 0.05	
\\ A3391 & 750 & 300 & 0.8 & 0.08	
\\ A576 &  700 & 150 & 0.6 & 0.06	
\\ PKS 0745-191 & 1050 & 100 & 1.5 & 0.08	
\\ A644 & 860 & 200 & 0.6 & 0.08	
\\ A754 & 800 & 300 & 1.6 & 0.09   	
\\ HYD-A & 680 & 220 & 1.8 & 0.07
\\ A1060 & 500 & 150 & 0.7 & 0.09
\\ A1367 & 700 & 500 & 0.4 & 0.03	
\\ VIRGO & 600 & 90 &  1.0 & 0.02 	
\\ CENT & 400 & 120 & 1.6 & 0.06	
\\ COMA & 1000 & 500 & $-$ & $-$ 	
\\ A1644 & 700 & 180 & 1.8 & 0.08  	
\\ A3532 & 660 & 350 & 0.8 & 0.12          
\\ A1650 & 600 & 100 & 0.8 & 0.08 
\\ A1651 & 900 & 220 & 1.0 & 0.12 
\\ A1689 & 1150 & 200 & 1.4 & 0.15
\\ A1736 & 800 & 700 & 1.2 & 0.05
\\ A3558 & 970 & 600 & 2.0 & 0.15
\\ A3562 & 550 & 180 & 0.8 & 0.12
\\ A3571 & 850 & 220 & 1.1 & 0.08
\\ A1795 & 710 & 70 & 0.5 & 0.07
\\ A2029 & 920 & 100 & 0.8 & 0.06
\\ A2052 & 550 & 200 & 2.1 & 0.08
\\ MKW3s & 550 & 100 & 0.8 & 0.08
\\ A2065 & 750 & 200 & 2.1 & 0.10
\\ A2063 & 600 & 150 & 1.0 & 0.08
\\ A2142 & 1050 & 200 & 1.8 & 0.12  
\\ A2199 & 670 & 150 & 1.3 & 0.08
\\ A2204 & 1100 & 80 & 1.0 & 0.08  
\\ TRI AUST & 800 & 220 & 2.5 & 0.05
\\ A2244 & 930 & 150 & 0.5 & 0.05 
\\ A2256 & 920 & 550 & 0.9 & 0.15
\\ OPHI & 880 & 220 & 2.5 & 0.06 
\\ A2255 & 1000 & 650 & 0.5 & 0.07 
\\ A2319 & 870 & 180 & 0.4 & 0.08
\\ CYG-A & 850 & 50 & 2.5 & 0.05
\\ A3667 & 730 & 250 & 0.5 & 0.04
\\ A2597 & 850 & 100 & 1.6 & 0.10  
\\ KLEM44 & 550 & 100 & 0.6 & 0.08
\\ A4059 & 560 & 110 & 0.4 & 0.09

\end{tabular}
\end{center}
%
%
\end{table}

\subsection{X-ray Luminosity}
\label{xray:luminosity}

In Fig. \ref{luminosity} we plot the bolometric  
X-ray luminosity against total mass deposition rate.
The bolometric luminosity was retrieved from the work of 
David et al. (1992) and the mass deposition rates come from our analysis.
We have used HRI observations here only when no PSPC observation was
available.
From this plot we notice that the upper left and the lower right corners are 
not populated. 
The former can be taken as an expression of the fact that for a given 
X-ray luminosity there is always a maximal mass deposition rate.
(Note that the cooling luminosity $L_{\rm cool}\sim \dot{M}  \times T$.)
The latter gives us a clear indication that very luminous systems harbour massive flows.



\begin{figure}
\centerline{
\psfig{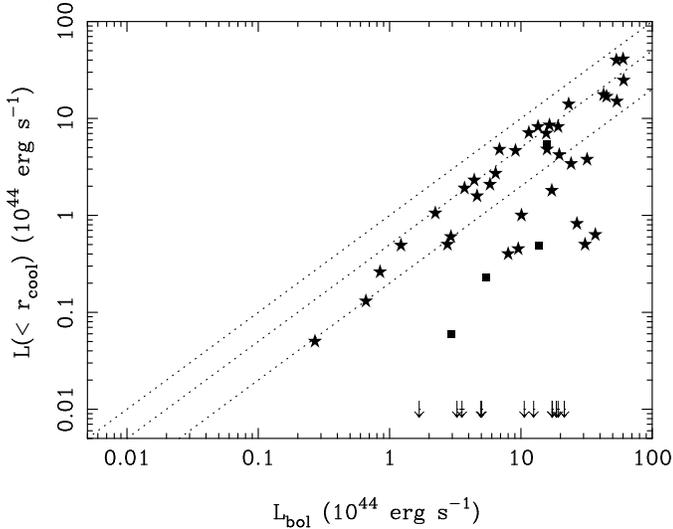}
}
\caption{
The luminosity from the cooling region,  $L(<{r}_{\rm cool})$, is plotted against the 
X-ray bolometric luminosity from David et al. (1992). We note that, despite the dispersion, 
many clusters have $L(<{r}_{\rm cool})$ contributing to 50 percent or more to the 
total X-ray luminosity. The dotted lines represent 20, 50 and 100 percent of the 
cluster X-ray bolometric luminosity. Arrows represent non-cooling-flow clusters which were
assigned  $L(<{r}_{\rm cool})$=0.01 for display purposes only.
}
\label{frac-luminosity}
\end{figure}

We have also investigated how much of the total bolometric luminosity 
can be attributed to the flow for each cluster in the sample. To answer 
this question we computed the luminosity from 
within the cooling radius,  $L(<{r}_{\rm cool})$
\footnote{Note that  $L(<{r}_{\rm cool})$ 
includes direct cooling
and the gravitational work done on the gas; direct cooling accounts for only about 
half of $L(<{r}_{\rm cool})$.},
and plotted  it against the bolometric X-ray luminosity given by David et al. (1992) in 
Fig. \ref{frac-luminosity}.
We find that, despite the spread, 
20 percent of the clusters have 50 percent or more of 
their bolometric X-ray luminosity produced in a region $<{r}_{\rm cool}$.
The cooling flow region in A2204 contributes  
more than 70 percent to the cluster's total X-ray luminosity. The percentages for 
the other clusters are listed in Table \ref{results}. 

\begin{figure}
\centerline{
\psfig{figure=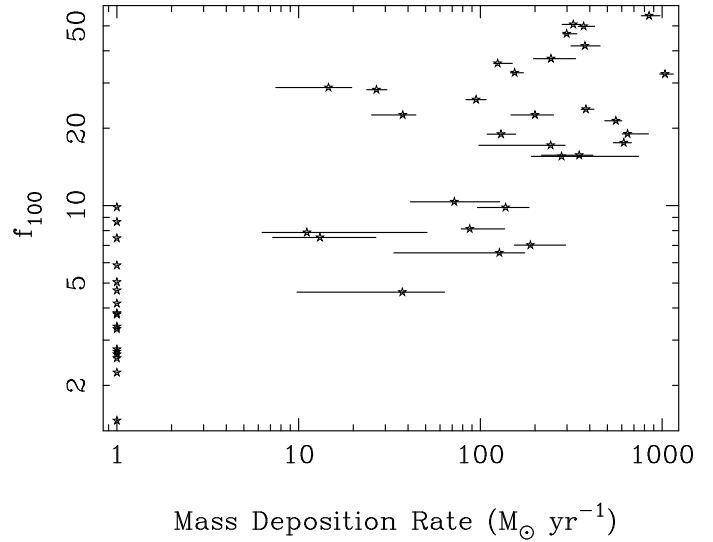,width=0.6\textwidth,angle=270}
}
\caption{
$f_{100}$ is defined as the luminosity from a region of 100 kpc radius 
divided by the cluster X-ray bolometric luminosity given by  David et al. (1992).
We note that non-cooling-flows (displayed here with $\dot{M}$ = $1 {\rm M}_{\odot}$ yr$^{-1}$)
have $f_{100} < 10$ percent, whereas cooling flow clusters have values of 
$f_{100}$ ranging from 5 to 50 percent. A trend of increasing $f_{100}$ with mass deposition 
rate is indicated  by the data.
}
\label{lum100}
\end{figure}

The approach described in the last paragraph does not treat non-cooling-flows 
properly since the concept of a cooling radius, ${r}_{\rm cool}$, in this case is meaningless.
To compare cooling flows and non-cooling-flow clusters  we have computed 
the fraction of the total luminosity coming from a central region of 100 kpc radius.
This quantity, $f_{100}$, is plotted against $\dot{M}$ in Fig. \ref{lum100}
from which is clear that:   
(i) non-cooling-flows have an overall smaller 
fraction of their X-ray luminosity emitted from the centre ($f_{100} \sim 1-10$ percent), 
when compared to cooling-flows ($f_{100} \sim 5-50$ percent; clusters 
with massive cooling flows show  $f_{100} > 15$ percent) and (ii) a trend of 
increasing $f_{100}$ with  $\dot{M}$ is indicated by the plot 
Some of these issues were investigated previously by Edge (1987) and 
White et al. (1997), but to our knowledge, it is the first systematic investigation 
on the fraction of a cluster luminosity contributed by its core, undertaken with
{\it ROSAT}. 

We note that the large contribution to the X-ray bolometric luminosity emerging from  
within the inner $\sim$200 kpc of a cooling flow cluster must have an influence 
on its detectability in imaging surveys (Pesce et al. 1990) and on the scatter  
in the $L_{\rm x} - T_{\rm x}$ correlation (Fabian et al. 1994).


\subsection{New Cooling Flows in the Sample}
\label{new:cfs}

\begin{figure*}
\centerline{\psfig{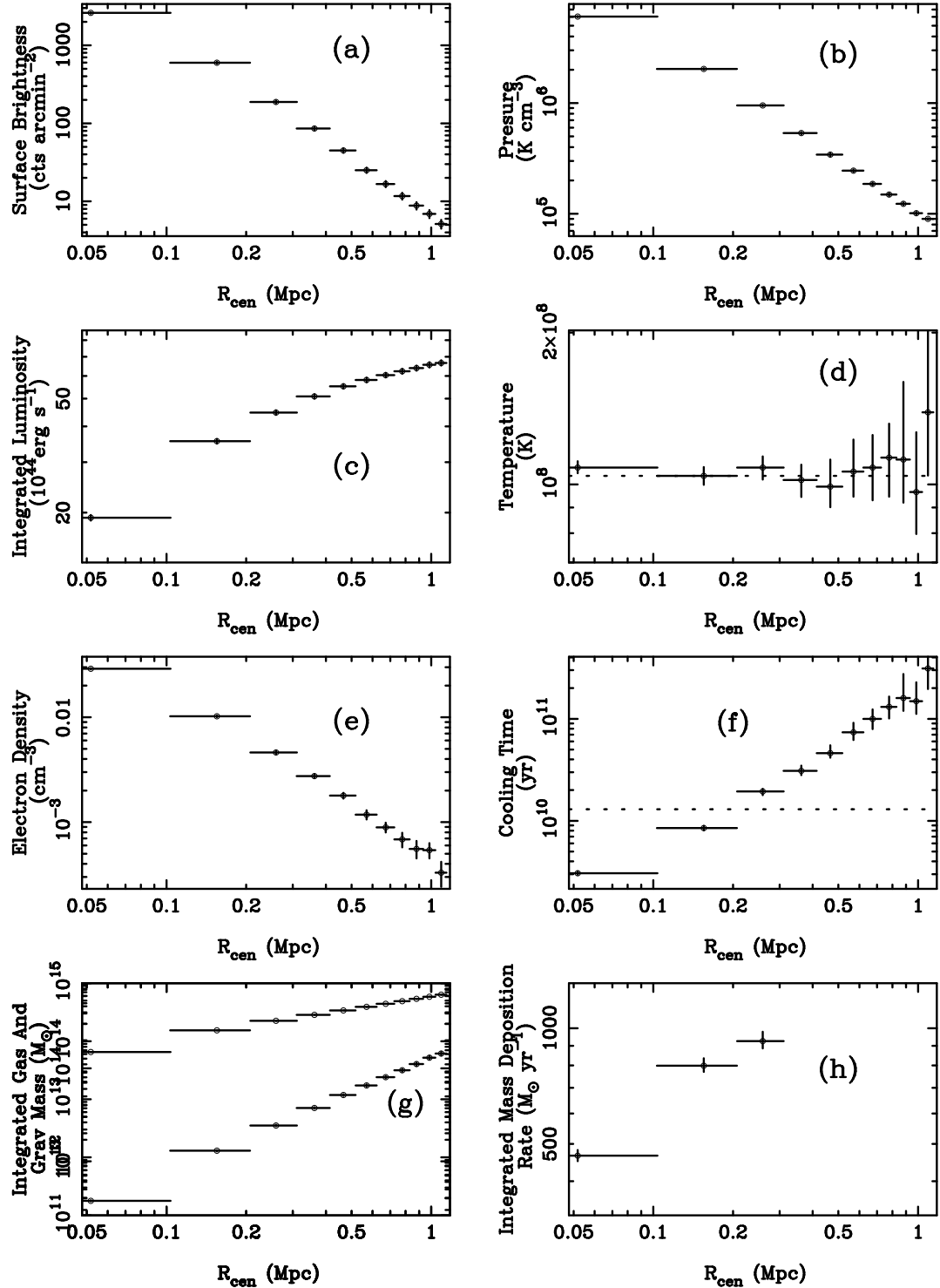}}
\caption{Summary of the deprojection results for the PSPC observation of
the cluster A2204. Points in Figs (a), (b), (c), (e) and (g) represent 
the mean value and 1$\sigma$ errors (in each radial bin) from 
100 Monte-Carlo simulations. Points in Figs (d), (f) and (h) represent
the median and 10 and 90 percentile estimations from 
100 Monte-Carlo simulations. The dotted line in panel (d) marks the average 
temperature from David et al. (1992), and $R_{\rm cen}$ denotes the radius at the bin centre.
Note that the potential is adjusted 
such as to produce a flat temperature profile. 
}
\label{a2204}
\end{figure*}

We present in Table \ref{newcflows} 
the mass deposition rate, cooling radius, and
central cooling time for the clusters which did not feature this
information in the Edge et al. (1992) paper. 
Three interesting cases are A2065, A1650 and A2204. In the first two cases
the resolution of the data in the Edge et al. paper  
hid a modest and a large cooling flow. The third case (A2204) is a  
massive cooling flow.
A summary of the deprojection results for this cluster 
is presented in Fig. \ref{a2204}. (A2204 was observed with the ROSAT PSPC 
from 1992 September 04 to 1992 September 05 
and with the ROSAT HRI from 1995 January 01 to 1995 January 11, with total exposure times
of 5.36 ks  and 15.5 ks respectively.)

\begin{table}
\caption{{\bf New cooling flows in the B55 Sample.}
The values quoted for $\dot {M}$, $r_{\rm cool}$, and 
$t_{\rm cool}$ are median, 10 and 90 percentile estimates from the
deprojection algorithm. The units used for $\dot {M}$, $r_{\rm cool}$,
and $t_{\rm cool}$ are M$_{\odot}$ yr$^{-1}$, kpc, and Gyr respectively.}
\label{newcflows}
\begin{center}
\begin{tabular}{ccccccc}
Cluster & $z$ & $\dot {M}$ & $r_{\rm cool}$ & $t_{\rm cool}$ \\
\hline
\\
Klem44 & 0.0283                     & 87$^{+25}_{-19}$   &\
         133$^{+42}_{-27}$  & 2.3$^{+0.6}_{-0.3}$    \\
\\
A1650  & 0.0845                     & 280$^{+464}_{-89}$ &\
         165$^{+103}_{-24}$  & 2.4$^{+1.2}_{-0.8}$  \\
\\
A1651  & 0.0846                     & 138$^{+48}_{-41}$  &\
         127$^{+32}_{-31}$  & 6.5$^{+0.7}_{-0.7}$  \\
\\
A2204  & 0.1523                     & 852$^{+127}_{-82}$  &\  
         199$^{+60}_{-44}$  & 3.1$^{+0.1}_{-0.1}$  \\
\\
A2065  & 0.0722                     & 13$^{+14}_{-6}$  &\  
         56$^{+22}_{-23}$  & 4.4$^{+2.2}_{-1.3}$  \\
\\
A3558  & 0.0478                     & 40$^{+39}_{-10}$  &\  
         68$^{+75}_{-20}$  & 10.2$^{+0.3}_{-0.2}$  \\

\end{tabular}
\end{center}
\end{table}

At the same time that new cooling flows were discovered three clusters considered to 
have small flows are now classified as non cooling flows: A2319, A576, and A754. 
A2319 does not have the single  peak typical of a cooling flow when observed with 
the HRI. The central cooling time is marginally consistent with the assumed age, 
and it posseses a radio halo (Hanisch 1982). 
A754 has a very disturbed X-ray morphology and a central cooling time in excess of 
13 Gyr. It is known to be in a merging state (Henry \& Briel 1995, Zabludoff \& Zaritsky 1995,
and Henriksen \& Markevitch 1996). 
A576 has a very flat surface 
brightness in X-rays when observed with the HRI. Although the cooling time is low
the mass deposition profile is perturbed and the value of the 
mass deposition rate is consistent with zero. 
From our analysis a clear cooling-flow/morphology relation appears
in the B55 sample. This is consistent with the results of 
Buote and Tsai (1996) who found a quantitative correlation between 
the mass flow rate and morphology of 37 ROSAT PSPC clusters 
of the B55 sample. A similar analysis for the whole B55 sample 
will appear elsewhere (Peres, Buote \& Fabian, in preparation.)

All clusters which we classify as non-cooling-flows 
display signs of merger activity and/or a disturbed morphology in X-rays.
Most notably, there are no examples, in the B55 sample, 
of a cooling flow cluster without a well defined central galaxy 
at the bottom of the cluster potential well~\footnote{  
The existence of non-cooling-flow clusters with a cD galaxy at the 
bottom of the potential well, e.g. A399 and A401,
is very rare leading to the interpretation of it  
as evidence of a previous phenomenum, which 
disrupted the flow (McGlynn \& Fabian 1984; see
Fabian, Peres \& White 1997 for the specific 
case of A399/A401).}.  
The lack of ROSAT data for A2147 made us keep its classification as 
a non-cooling-flow cluster,  since it had a large central
cooling time and a mass deposition rate consistent with zero in the analyses 
of Edge et al. (1992) and White et al. (1997).


\subsection{Fraction of Cooling Flows}
\label{fraction}

The  work of Edge et al. (1992) addressed this issue within the
B55 sample, concluding that the fraction of
cooling flow clusters  is 70 percent. However they warned that
it could be ``{\it 90 percent, once the
effect of different bin sizes is taken into account}''. 
In the study by Edge et al. (1992) only 36 clusters 
(65 percent) had observations with
a resolution better than 100 kpc.
Correction for that was made on the basis of a
tendency for the cooling time to decrease with bin size as 
$t_{\rm c} \sim R^{1.5}$, 
which the authors inferred from their sample.

\begin{figure}
\centerline{
\psfig{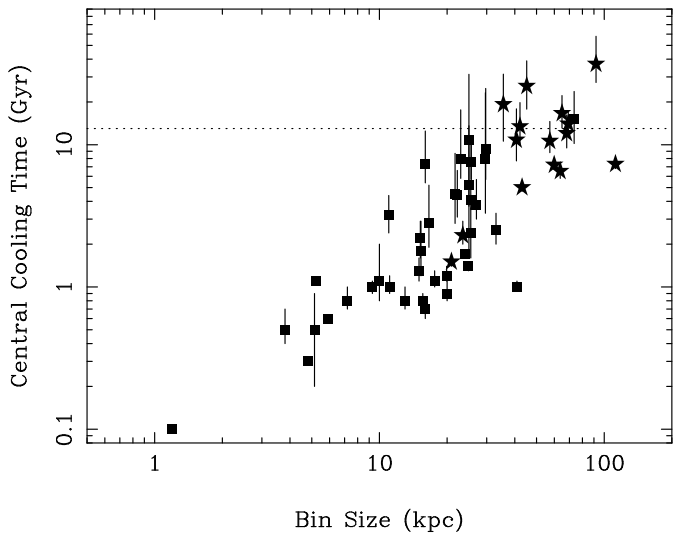}
}
\caption{
Central cooling time plotted against binsize 
for the B55 sample. The dotted line indicates the assumed age of the
Universe (13 Gyr). From this plot it is evident that 70-90 percent
of the clusters in this sample are cooling flows. The trend of reduced 
central cooling time with increased resolution is also easily observed for
the sample as a whole. The stars and squares represent PSPC and HRI observations 
respectively.}
\label{binsize}
\end{figure}

\begin{figure}
\centerline{
\psfig{figure=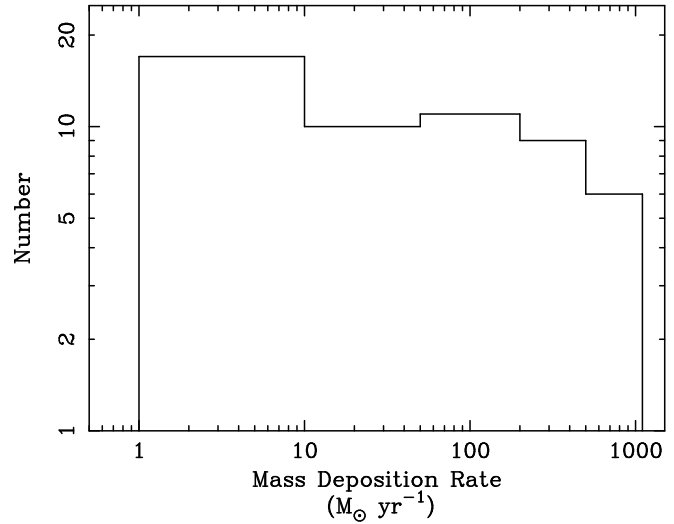,width=0.6\textwidth,angle=270}
}
\centerline{
\psfig{figure=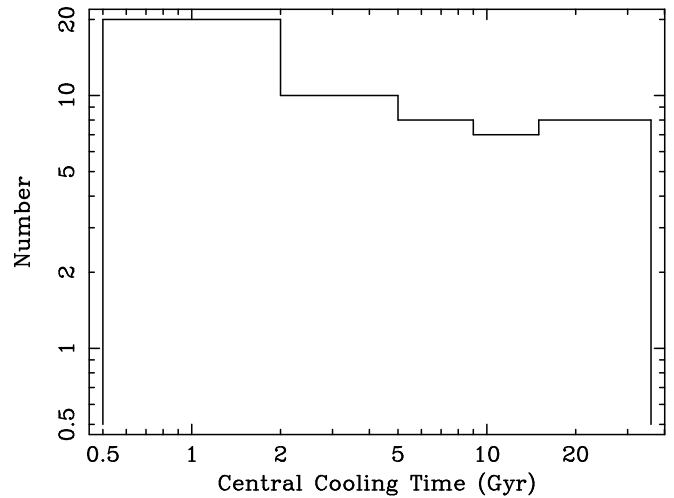,width=0.6\textwidth,angle=270}
}
\caption{Histograms of mass deposition rates (top) and central cooling
times (bottom) for the ROSAT observations of the B55 clusters. Note
that changing the value of {\it H}$_0$ does not affect the fraction of
cooling flows in the sample, because the ratio 
t$_c$/{\it H}$_0^{-1}$ scales as  {\it H}$_0^{0.5}$ (cf. Fabian et al
1984).}
\label{hist_tc_mdot}
\end{figure}

Here we confirm this tendency (cf. Fig. \ref{binsize}), 
but do not resort to it to establish the 
fraction of cooling flows in the B55 sample.
With the improved spatial resolution of ROSAT 
we can resolve the central cooling regions of all clusters in the B55
sample as can be seen from the short cooling times displayed in the
histogram in Fig. \ref{hist_tc_mdot}. 


We define here a cluster as a non-cooling-flow when the upper limit (90th percentile)
to the central cooling time, as given by the deprojection 
procedure, is larger than the assumed age of the cluster
(cf. Sec. \ref{method} and Figs. \ref{binsize}, \ref{a2204} above).  
By doing so we conclude that the fraction of cooling flows in the B55
sample is 70 percent 
(consistent with the 70-90 percent estimates of
Edge et al. 1992). Importantly, this is a conservative value since  
we are not considering as cooling flows clusters for which 
the lower limit to the central
cooling time
is consistent with their assumed age (i.e. 13 Gyr).
If we did include the clusters for which the central cooling 
time is consistent with 13 Gyr then the fraction of flows 
would be $\sim$90 percent and we would have to include as cooling-flows
%
%
A119, A401, A399, A3158, A3266, A754, A3532, A1736, Triangulum Australis, 
A2256, A2319, and A3667. 
We prefer the first possibility here because it does not include as
cooling flow those clusters which are clear mergers, but whose  
central cooling time is just above the assumed age of the Universe;
the ages of these clusters are certainly 
less than the latter.

In hierarchical scenarios for the 
formation of structures in the Universe clusters are assembled from smaller
subunits through mergers. 
This complicates the estimation of the age of a cluster (and of its cooling flow age), which is one of  
the parameters used to determine the mass deposition rate. 
The fact that a cluster has suffered a merger does not 
necessarily imply that its cooling flow has an age equal to the lookback time 
of the merger, i.e. cooling flows may be older than their recently formed 
host cluster; as yet we do not know in detail how a merger affects  
an existing cooling flow
\footnote{A85 is the example of a cluster with clear substructure which harbours a strong 
cooling flow.}. 
In fact if only approximately equal-mass, 
head-on collisions disrupt existing cooling flows 
then the rarity of such events is reconciled with the high fraction of flows in 
our sample and can be used in future larger datasets to constrain cosmological scenarios.

A detailed investigation of the sensitivity of the  mass deposition rate and 
fraction of cooling flows to individual cluster age would be ideal, but the complex history  
of formation of a cluster justifies the use of an average cluster age in our analysis.
Assuming  an average age of 13 Gyr for all clusters we obtained a fraction of cooling flows 
of 70-90 percent in the B55 sample.  
Since clusters are younger than 13 Gyr in all plausible formation scenarios
$\dot{M}$ and the fraction of cooling flows 
derived here are upper limits to their true value.
Assuming an average cluster age of 6 Gyr yields a fraction of 
cooling flows equal to 65 percent, whereas an age of 2 Gyr would still classify 
45 percent of our clusters as cooling flows.
The values of $\dot{M}$ do not change by more than a
factor of three when we assume an average cluster age of 5 Gyr (With the exception of
A644, A1650, A1651, A2142 and A4059, they either change by less than a factor of two or 
do not change appreciably.)  

The high fraction of cooling flows 
found here is objectively defined and secure. 
The sample is not biased by the presence 
of large cooling flows because it is flux-limited and constructed from 
observations with broad-beam instruments.
All the clusters considered as cooling flows have very short 
central cooling times, and the fraction of flows in the sample 
remains basically unaltered if any
other (reasonable) cosmology is selected (t$_c$/{\it H}$_0^{-1}$ scales as  {\it H}$_0^{0.5}$; 
cf. Fabian et al 1984).

\begin{figure}
\centerline{
\psfig{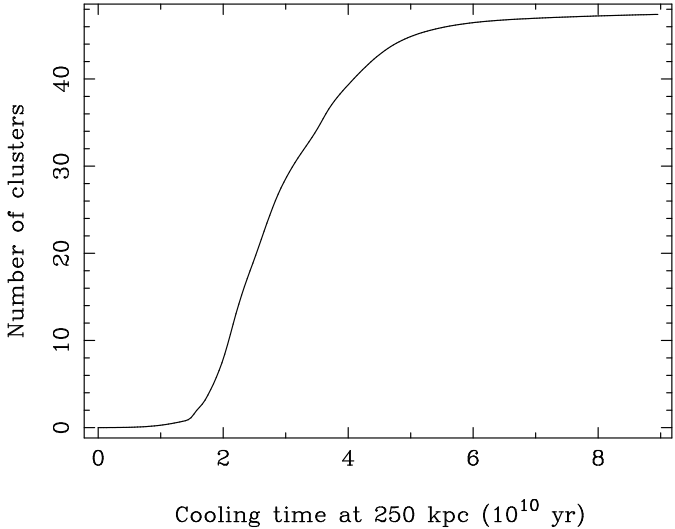}
}
\caption{
Cooling time at 250 kpc plotted against cumulative number of clusters. The plot shows 48
clusters for which the deprojection analysis was undertaken to a radius greater than 
250 kpc. 
Each cluster was represented by a normalized gaussian with 
dispersion given approximatelly by the deprojection limits.
This plot shows that cooling times at 250 kpc from the centre of 
cooling-flow clusters are higher than 
the age of the Universe (13 Gyr) by not more than a factor of 
$\sim$4.}
\label{tc250}
\end{figure}

We have also computed the cooling time of the hot phase
at a fixed radius (250 kpc) for as many clusters in our sample as possible.
The value quoted  in Table \ref{results}, ${t}_{250}$, is the cooling time of 
the bin encompassing the 250 kpc radius and the quoted limits come 
from the Monte-Carlo implementation of the deprojection analysis.
The results are summarised in Fig. \ref{tc250}, from which  is clear that 
all but two clusters have cooling times $13 < {t}_{\rm cool}(250 \, {\rm kpc}) < 51$ Gyr. 
Fig. \ref{tc250} was produced under the approximation that the 
limits to ${t}_{250}$ were 1$\sigma$ extrema of a gaussian distribution
(the limits were computed from the uncertainties provided by the deprojection analysis). 

This exemplifies once more the need for high spatial resolution in the 
determination of a cooling flow and provides a first estimate of the range of 
cooling times at around 250 kpc required  
from future numerical simulations of clusters of galaxies. The short cooling 
times at 250 kpc found in our sample are at odds with the values in 
the simulations by Cen et al. (1995) and support the multiphase picture of 
the ICM where there can be gas cooling at and beyond 250 kpc. 
Future higher resolution data may demonstrate that some of the the B55 clusters   
which have central cooling times larger than their assumed agee (13 Gyr here) 
are nonetheless cooling-flows,
as indicated from the trend in Fig. \ref{binsize} (although the trend is not guaranteed for 
every cluster individually, meaning that there are clusters for which we do not expected 
to detect a cooling flow even with data of improved resolution).


\begin{figure}
\centerline{
\psfig{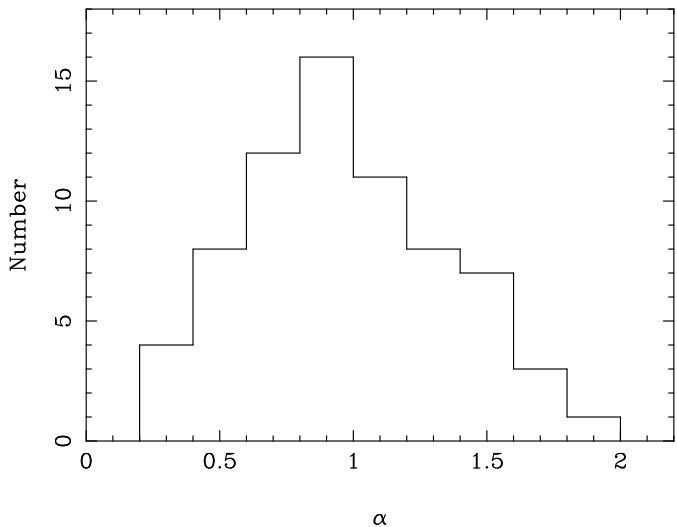}
}
\caption{Histogram with the values of the slope of the integrated 
mass deposition profile for the B55 clusters. Note that the values
scatter around the value $\alpha \sim 1$ predicted by 
multiphase models
.}
\label{hist_alpha}
\end{figure}


\subsection{Mass Deposition in Cooling Flows}
\label{profiles}

Cooling flows are known to deposit large amounts of cool material 
throughout a cluster core (cf. Sec. \ref{fraction} and Fig. \ref{hist_tc_mdot}). 
From table \ref{results} is clear that the accreted mass during the 
age of the flow, $M_{\rm acc} \sim \dot{M} \times t_{\rm age}$, varies 
from $6 \times 10^{10} \Msun$ to $6 \times 10^{12} \Msun$, if we assume an
average cluster age of 6 Gyr and mass deposition rates of 10 and 1000 \Msunpyr
respectively.
We note from the discussion in the previous section that these estimates are not 
altered by more than a factor of $\sim 6$ when the cluster age is halved since  
$\dot{M}$ does not change by more than a factor of $\sim 3$. The mass accreted 
in large cooling flows ($> 200 \Msunpyr$) should not change by more than $\sim 2$.
Therefore it would be desirable to understand how this material is distributed 
over the cooling region, and how the cooling cores are distributed in the Universe.

We have studied the spatial variation of this material 
over the cooling regions of the B55 clusters with the HRI.
To avoid contamination by small scale substructure 
we use the integrated mass deposition (IMD) profile which comes from the deprojection analysis 
and therefore has information on the azimuthally averaged mass
deposition only.
The shape of the IMD profile can be used to test the 
multiphase nature of the ICM and to trace the evolutionary 
history of the  flow.  

As for the first case, we know from previous work (Nulsen 1987, 
Thomas et al. 1987) that simple models of 
a multiphase flow would lead to an IMD profile of the form 
$\dot {M} (<{r}) \sim {r} ^{\alpha}$, with $\alpha \sim 1$.
We have used HRI and some PSPC observations to compute 
the slope of the IMD profile.
For this purpose we have obtained the profiles from the deprojection algorithm 
with data points given by the mean and 1$\sigma$ deviations. 
The profiles were then fit with a power law model. 
Only clusters with 4 or more bins inside the cooling radius
were considered in the fitting procedure.
Averaging the results for the whole set of observation analyzed here
yields $\langle \alpha \rangle = 0.95$ and a histogram of the values obtained 
can be seen in Fig. \ref{hist_alpha}.
The data is in good agreement with the simple multiphase models for the ICM.
(Arnaud 1987  reached  similar conclusions from a different sample.)

\begin{figure}
\centerline{
\psfig{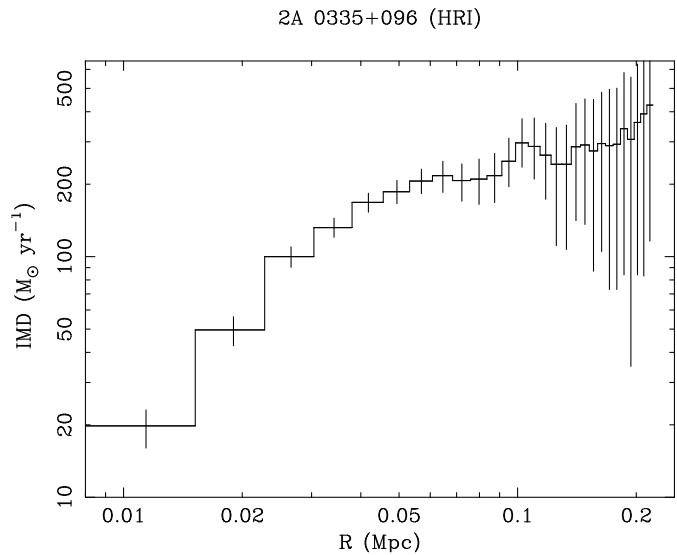}
}
\caption{
Integrated mass deposition (IMD) profile from the deprojection of an HRI 
observation of 2A 0335+096. Note that the slope changes clearly inside the cooling radius 
for the cluster (r$_{\rm cool}$ $\sim$ 200 kpc). Error bars are 10th and 90th percentiles
from our deprojection analysis.}
\label{2a03-IMD}
\end{figure}

The possibility of using the IMD profiles as tracers of the evolutionary history 
of the flows was raised by Edge et al. (1992). They searched for the  
existence of `breaks' or `plateaus'  in the IMD profiles, as indicative of previous 
phenomena which perturbed the flow.  
The detection of such features in the IMD profiles depends on how well 
we can sample the cooling region, for which HRI data is ideal.
To search for a `break' in the IMD profiles we fit them with a broken power 
law in the same manner described above. 
We have noticed that many clusters show some kind of break in their 
IMD profiles but only few are as apparent as in Fig. \ref{2a03-IMD} 
(see also Irwin \& Sarazin, 1996); the clusters showing clear breaks in our analysis
are 2A0335+096, A426, and A496.  
%
%
In the case of 2A 0335+096 the cooling time at around 70 kpc, where the break occurs, 
is $\sim$ 4-5 Gyr; in the suggestion by Edge et al. (1992) this is  
the approximate age of the flow (time since last disruption).
Some of the breaks, however happen in the outer bins of the IMD profiles, 
which can be attributed to 
fluctuations in the temperature profile, since $\dot{\rm M}$ goes to first order as 
$L_{\rm X}/T$. Allen and Fabian (1997) have proposed a method to estimate 
the age of cooling flows based on the spectral capabilities of the 
PSPC which needs, however, high signal-to-noise data. 
Future high spatial resolution missions like {\it AXAF} will pin-down 
this issue. 
\footnote{Numerical experiments are currently being undertaken with the aim of 
understanding how cluster collisions can affect the central cooling flow
(e.g. Burns et al. 1994, Roettiger et al. 1997).
Once the cause of disruption of a cooling flow is known, 
the incidence of cooling flows (sec. \ref{fraction}) 
may be used to constrain cosmological scenarios.}


\begin{figure}
\centerline{
\psfig{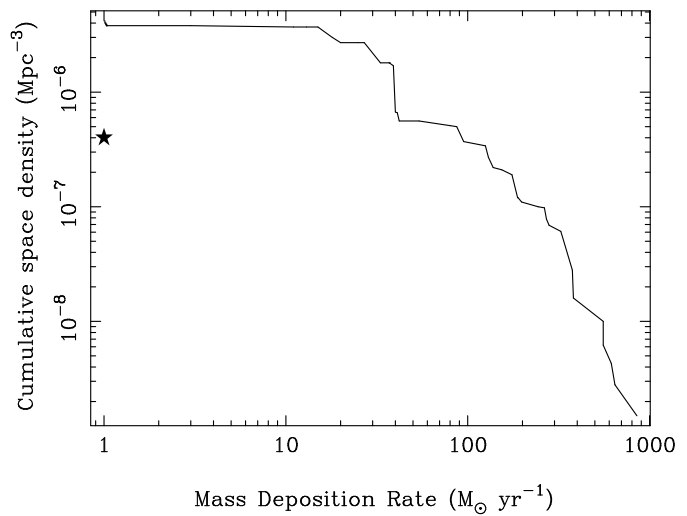}
}
\caption{
Cumulative space densitive of cooling flows from the B55 sample.
The density is corrected for the selection function $|b| >20$ degrees. 
The space density of non-cooling-flows, which were assigned here a mass deposition rate
equal to one, is represented by a star.
}
\label{space-density}
\end{figure}

We have also used our recent data on the mass deposition by cooling flows to recalculate 
the space density of the flows (Edge et al. 1992). We have selected the clusters with 
galactic latitude larger than 20 degrees and applied the usual procedure to compute
the density function for a flux-limited sample. The result is presented in Fig. 
\ref{space-density}; the slope of a power law fitted between 
$\dot M = 10$ \Msunpyr and $\dot M = 1000$ \Msunpyr
is $\sim -$1.8. Selection of clusters with $|b| >10$, instead 
of $|b| >20$ does not change the results significantly.


\subsection{Gas fractions in cluster cores}
\label{baryon:frac}

Gas masses within 500 kpc or 250 kpc 
for the B55 sample are presented in Table \ref{results}. 
They range from 2.5 to 64$\times 10^{12}$ \Msun, constituting 
from 7 to 23 percent of the total gravitational mass within this region. 

Systematic errors due to our choice of potential and methodology 
are larger than the limits quoted from the deprojection analysis.
However, since the baryon fraction in clusters ($f_{\rm b}$) tend to increase 
with radius, we note that for our sample, 
$f_{\rm b}>6$ percent. This conclusion is robust to 
the systematics mentioned above,
agreeing with previously published results (White et al. 1993, and 
White \& Fabian 1995). 
It is important to stress that the gas fractions presented here do not include the contribution
of baryonic material in cluster galaxies, representing lower limits to the baryon fractions 
in clusters.

\begin{figure}
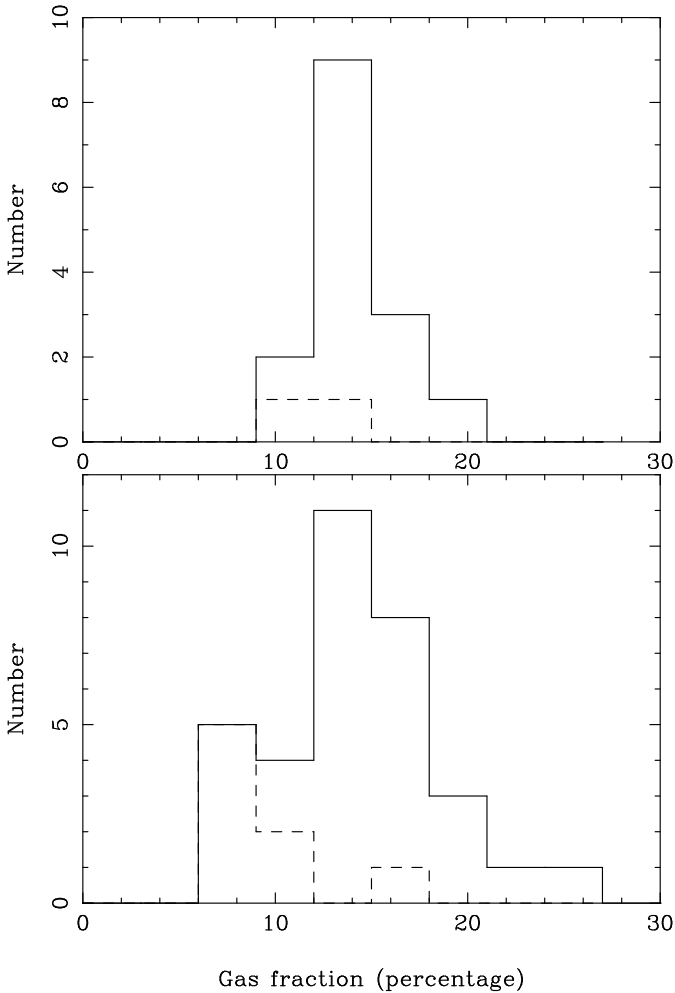

\centerline{
\psfig{figure=frac_non-cf.ps,width=0.5\textwidth,angle=270}
}
\centerline{
\psfig{figure=frac_cf.ps,width=0.5\textwidth,angle=270}
}
\caption{
The histograms above show the baryon fractions for non-cooling-flow clusters (top)
and for cooling-flow clusters (bottom). The dotted line in both panels represents the 
cases for which baryon fractions were calculated within 250 kpc only, whereas the solid line 
includes clusters with gas fractions computed within 500 kpc.
}
\label{baryon}
\end{figure}

We plot in Fig. \ref{baryon} the gas fraction for cooling-flows and non-cooling-flows
within 500 kpc from our analysis (250 kpc when information for 500 kpc is not available).
This plot indicates that the non-cooling-flows have a similar amount of gas, 
for a given mass, to cooling-flow clusters; the difference 
between them  residing therefore in the 
spatial distribution of the gas in the core.


\subsection{Comparison to Optical and Radio Data}
\label{optical/radio}

Cooling flows are the only simple route to
explain the wealth of X-ray data on the centres of clusters of galaxies.
From our previous analysis it becomes clear that: (i) cooling flows
are common and long-lived (fraction of at least 70 percent in a flux-limited
sample), and (ii) they deposit a large amount of cool material
throughout a cluster core (about 40 percent of the clusters in the
B55 sample have $\dot {\rm M}$ $>$ 100 \Msunpyr).

As a corollary of these points we state that the cooling flow plays an important
role on practically all phenomena happening in a cluster core.
Historically, two phenomena have been traditionally linked to the
cooling flow: optical emission line nebulosity and radio activity 
from the BCG. A review of the role of 
cooling flows in these phenomena can be found in Fabian (1994).

%
%
%
%

\begin{figure}
\centerline{
\psfig{figure=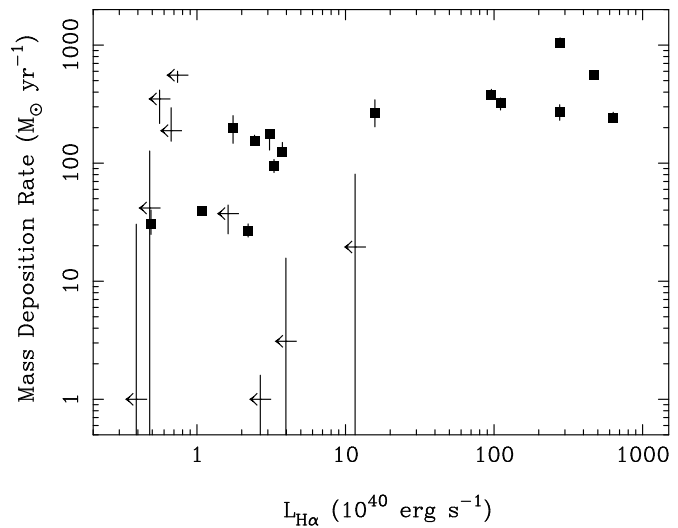,width=0.6\textwidth,angle=270}
}
\caption{Correlation of H${\alpha}$ luminosities from the centres of 
clusters listed in Heckman et al. (1989) and the mass deposition rates 
inferred from our deprojection analysis. Note that for clarity we did 
not include the Virgo cluster in this sample. Points with left-pointing
horizontal arrows represent upper limits for the observed
H${\alpha}$ luminosity.}
\label{halpha}
\end{figure}

\begin{figure}
\centerline{
\psfig{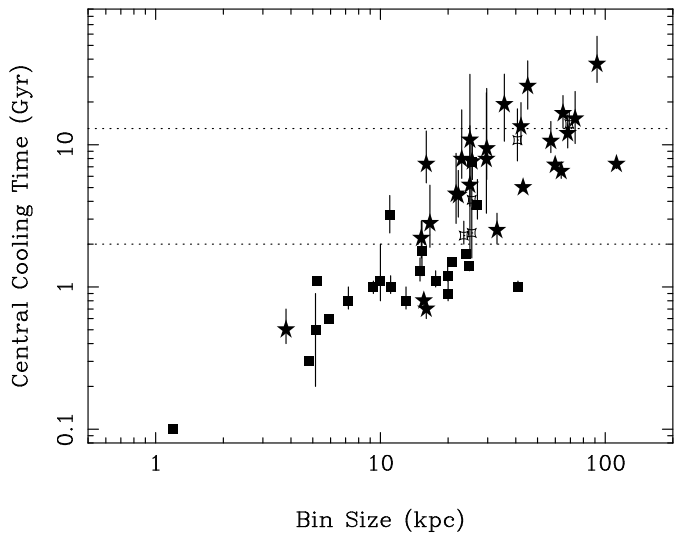}
}
\caption{In the diagram above we denote clusters with emission lines by
squares and 
clusters lacking emission lines by filled stars. Open squares are used
to represent the clusters for which we have no information on the presence of emission lines. 
Emission line clusters are situated predominantly below the line of  central cooling time equal to 2 Gyr. 
Note there are only two emission line cluster with central cooling times larger than 2 Gyr.}
\label{halpha_emislines}
\end{figure}

Here we use the data from the work of Heckman et al. (1989) to 
investigate the role of the flow in the  H${\alpha}$ luminosities
of the BCG. We have changed the values of the published luminosities to match 
our assumed cosmological parameters. 
The results of a cross-correlation between the B55 and Heckman's
sample (24 objects in  common) is displayed in Fig. \ref{halpha}.
We note that the weak trend suggested by Heckman et al. is maintained,
i.e. H$\alpha$ luminous galaxies lie in the centre of large cooling
flows although this special cluster environment does not guarantee
the existence of emission line nebulosity 
in its BCG.~\footnote{About 50 percent of the 
cooling flow clusters in the sample show 
line emission, whereas about 75 percent of
the clusters with emission lines display cooling flows with 
$\dot{M} > 100 \Msunpyr$. This trend can be used 
as an efficient method to 
search for large cooling flows at high $z$.}
Finally we note from Fig. \ref{halpha_emislines} that 
clusters which display emission line nebulosities 
have the shortest central cooling times in the sample: only two 
greater than 2 Gyr. 

Various generic scenarios have been put forward to understand the 
trend in Fig. \ref{halpha}. Crawford \& Fabian (1992) suggest that 
the emission line nebulosity is powered by mixing-layers in the multiphase,
turbulent ICM (Begelman \& Fabian, 1990). 
Since mixing layers occur where there is a population
of cold clouds and where the hot ICM is most turbulent, we would
expect emission-line nebulae to occur in the centres of large cooling
flows. On the other hand this scenario is consistent  with the observation that some 
large cooling flows do not host emission-line nebulae (e.g., A2029) since the hot ICM may be
in a quiescent state and/or most of the cold clouds may have formed stars
already. It is interesting to remember that A2029 shows 
little excess absorption (White et al. 1991, Allen \& Fabian 1997).
Allen (1995) has presented new evidence for star formation at the centres of 
cooling flows, proposing  that it can power the emission line nebulae; early suggestions 
of this mechanism are found in Johnstone et al. (1987),
Hu et al. (1985), and Heckman et al. (1989) among others. 
The role of magnetic reconection on the emission line nebulae 
have been re-investigated recently by Jafelice \& Friaca (1996), and a recent detailed 
case study has been presented by Voit \& Donahue (1997).

%
%

Optical images retrieved from the Space Telescope Digitized Sky Survey (DSS)
were used to locate the optical counterparts to the peak in the X-ray emission from 
our cluster images. The optical position listed in Table \ref{info} is 
the optical peak of the brightest galaxy nearest to the 
position of the X-ray peak. 
The listed offsets, $\Delta \theta$, are the modulus of 
the difference between the X-ray and optical positions, but no effort was made to correct 
the astrometry of the X-ray images. 
We note that most of the  clusters with large offsets, $\Delta \theta > 25$ arcsec, do not have
cooling flows: A119, A401, A3158, A754, A1367, A1736, A2256, and A3667. 
Most of the clusters, however have optical positions consistent with the uncertainty in 
the position of their X-ray peaks: 65 percent have $\Delta \theta \leq 10$ arcsec, and 
85 percent have  $\Delta \theta \leq 25$ arcsec.
The large values of $\Delta \theta$ for A1367 and A1736 stem from 
the flat X-ray surface brightness, without a clear centre. 
The extreme offset values  of A2256 and A754 are caused because the 
peak of the X-ray brightness does not correspond to the BCG in these 
merger systems, but to shocked regions of the ICM. 
The overall excellent agreement between the peak of the X-ray emission and 
the position of the BCG indicates that for most of the clusters 
in the B55 sample the X-ray gas is in a dynamically quiescent state, with the centre of 
the potential being occupied by a bright galaxy.

The appearance in the literature of a new, larger, photometrically 
homogeneous sample of BCGs (Lauer \& Postman 1994) has prompted us to use our 
new cooling flow results to search for a correlation between the 
strength of the flow and the optical light from the BCGs. This 
question has been investigated in the past in relation to the existence of a 
putative accretion population formed from the flow by  
Schombert (1987), Sarazin (1983), Thuan \& Puschell (1989), and Edge (1991).

\begin{figure}
\centerline{
\psfig{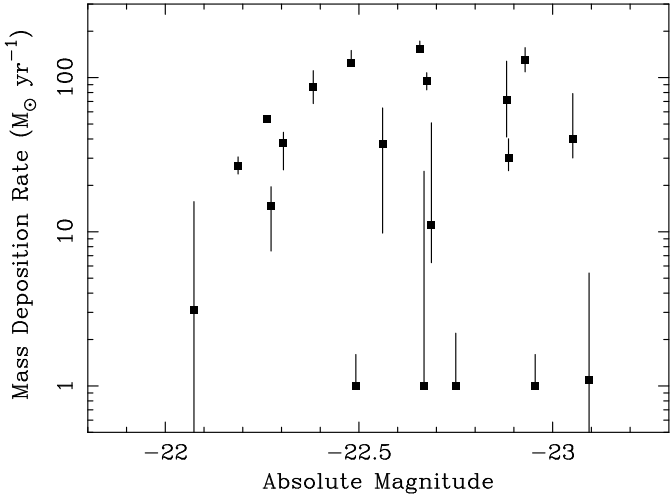}
}
\caption{The lack of correlation displayed  in the figure above indicates
that the putative accretion population from the flow is not composed of stars
with a normal IMF. This however does not eliminate the possibility of
ongoing low-mass star formation. Despite the lack of correlation, there is 
the indication of a cut-off in the upper left part of the plot. Non-cooling-flows were assigned 
$\dot{M} = 1$ for display reasons only.}
\label{lauer}
\end{figure}

The sample by Lauer \& Postman is volume-limited to c$z<$15,000 \kmps, 
and it has 19 BCGs (out of 119) in common with our sample. We plot 
$\dot {M}$ against the absolute magnitude in Fig. \ref{lauer},  
from which we notice that no evident correlation exists.
The lack of correlation can be understood if star formation 
is skewed towards low-mass objects in the environment of cooling-flows,
but this is a much harder question to answer observationally
(cf. Fabian 1994). 

\begin{figure}
\centerline{
\psfig{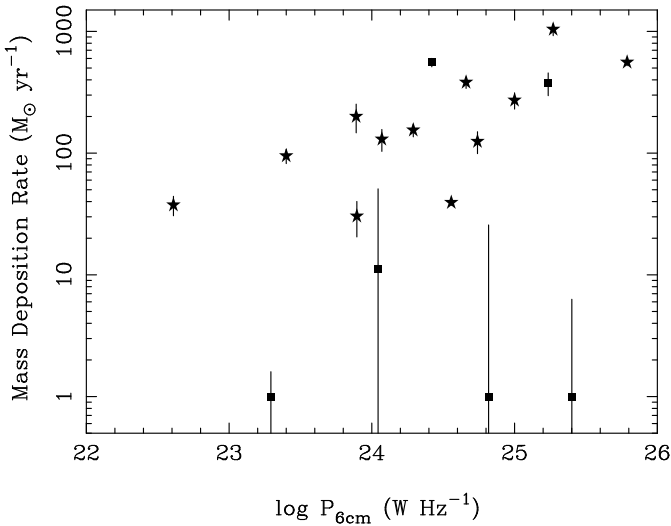}
}
\caption{The plot above shows a correlation between the 
integrated power at 6cm emitted 
by a cluster (P$_{6 {\rm cm}}$) and its cooling flow strength.
Points to the right
of the horizontal arrows represent upper limits for the observed
radio power. Stars represent  clusters with observed  optical 
line emission; squares are used to denote absence of emission lines 
or lack of information thereof. Non-cooling-flows were assigned $\dot{M} = 1$. 
Data for which only upper limits exist were not used in the plot.
}
\label{radio}
\end{figure}

We have also cross-correlated the strength of the flow with the 
radio power from BCGs. We have used radio data on  
17 BCGs by Ball et al. (1993) and archival data obtained from NED, 
and the NVSS survey. From the archival information we could 
obtain the flux density of the central source and then its luminosity at 
6-cm and/or 20-cm. 
No simple correlation is evident from Fig. \ref{radio}
(see also Fabian 1994), but as we  see, large cooling flows tend to 
have high radio luminosities (right upper corner of the diagram).
This trend can be understood as a consequence of the
high-pressure environment of cooling flows, which provide 
the working surface needed to produce the radio emission. 
When we split the data between clusters showing and clusters lacking emission lines,
we note that the trend exists for the former only. 
Inspection of Table \ref{info} shows that 19 out of the 22 clusters (i.e. 90 percent) 
which display emission lines have radio emission from the center, although 
only 19 out of 27 clusters (i.e. 70 percent) displaying central radio emission
have emission line nebulosity.

Finally we note a similarity between Figs. \ref{lauer} and \ref{radio}.
Both have a distinctive upper envelope with a roughly uniform distribution of data
under it. In neither figure do we see large cooling flows with 
a central cluster galaxy of 
small absolute magnitude or small radio power.


\section{Conclusions}
\label{conclusion}

Our analysis of the X-ray properties of cluster cores with {\it ROSAT}
showed that a cooling flow is the natural state of the cores of nearby 
clusters of galaxies. They constitute 70-90 percent of the clusters in 
our flux-limited sample and deposit more than  100 \Msunpyr
in about 40 percent of the clusters. Large  cooling flows can
contribute more than 70 percent of the cluster bolometric luminosity.

From our catalogue of ICM properties of the B55 sample we note
that cooling times at the central bin are small and cooling times 
at 250 kpc from the cluster centre do not exceed $\sim$4 times the 
age of the Universe (13 Gyr).  We show that 
mass deposition in cooling flows have an approximatelly 
linear dependence on radius and that 
clear breaks in some of the integrated mass deposition (IMD) profiles exist.
  
The cross-correlation of optical and radio properties with the strength
of the flow yielded weak trends only. We noted that overall large cooling 
flows tend to have luminous BCGs and powerful radio sources at their 
centres, whereas no such rule exists for non-cooling-flows.

%
%

\section*{ACKNOWLEDGMENTS} We are indebted to the members of the X-ray
group and to S. Sigurdsson for valuable discussions. 
We thank in particular C.S. Reynolds
and C.S. Crawford for providing us with some of the optical spectra
for the analysis of the optical emission lines. 
CBP thanks Conselho Nacional de Desenvolvimento 
Cientifico e Tecnologico (CNPq - Brazil) for financial support and
acknowledges the receipt of an Overseas Research Studentship (ORS) award. 
ACF, ACE and SWA thank the Royal Society for support.
DAW and RMJ thank PPARC for support.
\\
This research has made use of data obtained through the High Energy
Astrophysics Science Archive Research Center Online Service, provided by the
NASA-Goddard Space Flight Center, and the NASA/IPAC Extragalactic Database (NED)
which is operated by the Jet Propulsion Laboratory, California Institute of
Technology, under contract with the National Aeronautics and Space
Administration.
We also acknowledge the use of the online NVSS and FIRST catalogues.
We thank an anonymous referee for useful comments on the manuscript.

%
%
%
%
\clearpage
\begin{table*}
\caption{{\bf Summary of some deprojection results.}
(a) Name of the cluster. The letter inside brackets indicates wether 
the observation was made with the PSPC(P), or with the HRI(H); 
(b) Integrated mass deposition rate in units of \Msunpyr; 
(c) Cooling radius in units of kpc; 
(d) Central cooling time  in units of Gyr; 
(e) X-ray bolometric luminosity within the cooling radius,  L($<$ r$_{\rm cool}$), in units of 10$^{44}$\ergps; 
(f) Ratio of  $L(<{r}_{\rm cool}$) to $L_{\rm bol}$. The values of $L_{\rm bol}$ are from David et 
al. (1992);
(g) Cooling time of the bin encompassing the 250 kpc radius, in units of Gyr;
(h) Gas mass out to 0.5 Mpc or 0.25($^{\dagger}$) Mpc, in units of 10$^{12}$ \Msun; 
(i) Gravitational mass out to 0.5 Mpc or 0.25($^{\dagger}$) Mpc, in units of 10$^{12}$ \Msun; 
(j) Approximate ratio (percentage) of gas-to-gravitational mass inside 0.5 or 0.25($^{\dagger}$)  Mpc. 
}
\label{results}
\begin{center}
\begin{tabular}{cccccccccccccc}
Cluster & $\dot{M}$ & $r_{\rm cool}$ & $t_{\rm cool}$ & $L(<{r}_{\rm cool})$ & 
$\frac{{L}(<{r}_{\rm cool})}{{L}_{\rm bol}}$ & ${t}_{250}$ &
$M_{\rm gas}$ & $M_{\rm grav}$ & $\frac{{M}_{\rm gas}}{{M}_{\rm grav}}$ \\
\hline
(a) & (b) & (c) & (d) & (e) & (f) & (g) & (h) & (i) & (j) \\
\\ A85(H) & 107$^{+294}_{-31}$ & 93$^{+111}_{-10}$ & 1.0$^{+0.2}_{-0.1}$ & 2.7$^{+3.7}_{-0.2}$ & 
0.17$^{+0.23}_{-0.01}$ & $----$ & $----$ & $----$ & $----$  \\
\\ A85(P) & 198$^{+53}_{-52}$ & 146$^{+41}_{-41}$ & 2.4$^{+0.1}_{-0.1}$ & 4.8$^{+1.1}_{-1.2}$ & 
0.30$^{+0.07}_{-0.07}$ & 30.8$^{+2.3}_{-2.7}$ & 31.2$\pm$0.5 & 166 & 19$\pm$0.3 \\
\\ A119(P) & 0$^{+2}_{-0}$ & 0$^{+62}_{-0}$ & 19.2$^{+12.2}_{-8.6}$ & 0$^{+0}_{-0}$ & 
0 & 64.5$^{+2.0}_{-1.2}$ & 15.5$\pm$0.5 & 89 & 18$\pm$0.6 \\
\\ A262(P) & 27$^{+4}_{-3}$ & 104$^{+11}_{-10}$ & 1.5$^{+0.1}_{-0.1}$ & 0.3$^{+0.02}_{-0.02}$ & 
0.35$^{+0.02}_{-0.02}$ & 34.9$^{+9.9}_{-5.4}$ & 2.5$\pm$0.0$^{\dagger}$ & 33$^{\dagger}$ & 8$^{\dagger}$ \\
\\ AWM7(H) & 18$^{+123}_{-11}$ & 78$^{+91}_{-21}$ & 0.6$^{+0.2}_{-0.1}$ & 0.3$^{>0.7}_{-0.1}$ & 
0.11$^{+0.25}_{-0.04}$ & $----$ & $----$ & $----$ & $----$ \\
\\ AWM7(P) & 41$^{+6}_{-6}$ & 103$^{+5}_{-8}$ & 1.9$^{+0.2}_{-0.2}$ & 0.5$^{+0.03}_{-0.04}$ & 
0.18$^{+0.01}_{-0.01}$ & 28.0$^{+6.6}_{-3.2}$ & 15.4$\pm$0.2 & 89 & 17$\pm$0.2 \\
\\ A399(H) & 0.0$^{+51}_{-0}$ & 0$^{+110}_{-0}$ & 15.2$^{+8.4}_{-5.0}$ & 0$^{+0.7}_{-0}$ & 
0$^{+0.04}_{-0}$ & 35.1$^{+10.1}_{-7.2}$ & 22.2$\pm$1.5 & 163 & 14$\pm$0.9 \\
\\ A401(P) & 42$^{+82}_{-42}$ & 77$^{+66}_{-77}$ & 10.6$^{+4.0}_{-1.8}$ & 0.9$^{+1.6}_{-0.9}$ & 
0.03$^{+0.05}_{-0.03}$ & 22.5$^{+2.1}_{-1.7}$ & 35.3$\pm$0.7 & 235 & 15$\pm$0.3 \\
\\ A3112(P) & 376$^{+80}_{-61}$ & 192$^{+65}_{-48}$ & 1.9$^{+0.1}_{-0.1}$ & 7.0$^{+1.2}_{-1.2}$ & 
0.61$^{+0.10}_{-0.10}$ & 20.4$^{+2.2}_{-1.8}$ & 26.7$\pm$0.6 & 140 & 19$\pm$0.4 \\
\\ A3112(H) & 415$^{+252}_{-174}$ & 183$^{+38}_{-8}$ & 0.7$^{+0.1}_{-0.1}$ & 7.8$^{+0.3}_{-0.7}$ & 
0.68$^{+0.02}_{-0.06}$ & 12.5$^{+6.9}_{-2.0}$ & 10.1$\pm$0.3$^{\dagger}$ & 75$^{\dagger}$ & 13$\pm$0.4$^{\dagger}$ \\
\\ A426(H) & $>$427 & $>$78$^{}_{}$ & 0.6$^{+0.1}_{-0.1}$ & $>$6.8 & 
$>$0.29 & $----$ & $----$ & $----$ & $----$ \\
\\ A426(P) & 556$^{+33}_{-24}$ & 185$^{+11}_{-11}$ & 0.9$^{+0.0}_{-0.0}$ & 13.7$^{+0.4}_{-0.4}$ & 
0.59$^{+0.02}_{-0.02}$ & 20.5$^{+1.8}_{-1.7}$ & 34.4$\pm$0.4 & 130 & 26$\pm$0.3 \\
\\ 2A 0335+096(H) & 242$^{+234}_{-32}$ & 132$^{+93}_{-21}$ & 0.6$^{+0.4}_{-0.2}$ & 3.8$^{+1.3}_{-0.3}$ & 0.55$^{+0.19}_{-0.04}$ & $----$ & $----$ & $----$ & $----$ \\
\\ 2A 0335+096(P) & 325$^{+32}_{-43}$ & 215$^{+29}_{-29}$ & 0.9$^{+0.0}_{-0.0}$ & 5.0$^{+0.3}_{-0.3}$ & 0.73$^{+0.04}_{-0.04}$ & 19.2$^{+2.9}_{-3.2}$ & 18.5$\pm$0.5 & 86 & 22$\pm$0.6 \\
\\ A3158(P) & 25$^{+74}_{-25}$ & 65$^{+105}_{-65}$ & 12.0$^{+4.2}_{-2.5}$ & 0.4$^{+1.0}_{-0.4}$ & 
0.05$^{+0.12}_{-0.05}$ & 24.6$^{+2.8}_{-2.1}$ & 26.9$\pm$0.7 & 194 & 14$\pm$0.4 \\
\\ A478(H) & 520$^{+111}_{-107}$ & 192$^{+10}_{-5}$ & 1.1$^{+0.2}_{-0.1}$ & 14.5$^{+0.5}_{-0.3}$ & 
0.32$^{+0.01}_{-0.01}$ & 16.3$^{+3.8}_{-3.4}$ & 13.9$\pm$0.4$^{\dagger}$ & 108$^{\dagger}$ & 13$\pm$0.4$^{\dagger}$ \\
\\ A478(P) & 616$^{+63}_{-76}$ & 204$^{+27}_{-38}$ & 2.8$^{+0.1}_{-0.1}$ & 17.3$^{+1.7}_{-2.7}$ & 
0.38$^{+0.04}_{-0.06}$ & 15.3$^{+0.5}_{-0.5}$ & 44.1$\pm$0.5 & 272 & 16$\pm$0.2  \\
\\ A3266(P) & 0$^{+34}_{-0}$ & 0$^{+105}_{-0}$ & 13.7$^{+1.8}_{-1.1}$ & 0$^{+0.7}_{-0}$ & 
0$^{+0.04}_{-0}$ & 42.7$^{+3.9}_{-3.9}$ & 29.4$\pm$0.5 & 163 & 18$\pm$0.3 \\
\\ A3266(H) & 3.8$^{+35}_{-3.8}$ & 35$^{+78}_{-35}$ & 5.2$^{+8.4}_{-1.5}$ & 0.1$^{+0.03}_{-0.03}$ & 
0.5$^{+0}_{-0}$ & 43.3$^{+45.2}_{-5.2}$ & 22.1$\pm$1.6 & 146 & 15$\pm$1.1 \\
\\ A496(H) & 95$^{+37}_{-34}$ & 103$^{+16}_{-5}$ & 0.8$^{+0.2}_{-0.1}$ & 1.9$^{+0.2}_{-0.1}$ & 
0.32$^{+0.03}_{-0.02}$ & $----$ & $----$ & $----$ & $----$ \\
\\ A496(P) & 95$^{+13}_{-12}$ & 110$^{+12}_{-15}$ & 1.8$^{+0.1}_{-0.1}$ & 2.0$^{+0.1}_{-0.2}$ & 
0.33$^{+0.02}_{-0.03}$ & 38.4$^{+6.7}_{-6.1}$ & 6.5$\pm$0.1$^{\dagger}$ & 63$^{\dagger}$ & 10$\pm$0.2$^{\dagger}$ \\
\\ A3391(P) & 0$^{+5}_{-0}$ & 0$^{+32}_{-0}$ & 16.7$^{+5.6}_{-3.5}$ & 0.0$^{+0.1}_{-0}$ & 
0$^{+0.02}_{-0}$ & 37.0$^{+5.5}_{-3.1}$ & 16.1$\pm$0.5 & 161 & 10$\pm$0.3  \\
\\ A576(H) & 3$^{+13}_{-3}$ & 37$^{+38}_{-29}$ & 2.8$^{+2.4}_{-0.9}$ & 0.1$^{+0.1}_{-0.0}$ & 
0.03$^{+0.03}_{-0}$ & $----$ & $----$ & $----$ & $----$ \\
\\ PKS 0745-191(H) & 787$^{+368}_{-73}$ & 188$^{+23}_{-16}$ & 0.9$^{+0.2}_{-0.1}$ & 34.9$^{+1.8}_{-1.8}$ & 0.58$^{+0.03}_{-0.03}$ & 21.7$^{+19.8}_{-6.4}$ & 18.0$\pm$0.7$^{\dagger}$ & 149$^{\dagger}$ & 12$\pm$0.5$^{\dagger}$ \\
\\ PKS 0745-191(P) & 1038$^{+116}_{-68}$ & 214$^{+49}_{-25}$ & 2.2$^{+0.1}_{-0.1}$ & 43.6$^{+4.6}_{-2.8}$ & 0.73$^{+0.08}_{-0.05}$ & 18.3$^{+1.1}_{-1.2}$ & 55.7$\pm$1.1 & 329 & 17$\pm$0.3  \\
\\ A644(H) & 216$^{+48}_{-59}$ & 167$^{+18}_{-24}$ & 4.5$^{+4.2}_{-1.7}$ & 4.5$^{+0.6}_{-1.0}$ & 
0.23$^{+0.03}_{-0.05}$ & 25.6$^{+19.3}_{-6.2}$ & 9.3$\pm$0.4$^{\dagger}$ & 92$^{\dagger}$ & 10$\pm$0.4$^{\dagger}$  \\
\\ A644(P) & 189$^{+106}_{-35}$ & 141$^{+62}_{-18}$ & 6.8$^{+0.4}_{-0.4}$ & 4.1$^{+2.3}_{-0.6}$ & 
0.21$^{+0.12}_{-0.03}$ & 26.1$^{+1.4}_{-1.0}$ & 37.2$\pm$0.5 & 253 & 15$\pm$0.2 \\
\end{tabular}
\end{center}
\end{table*}
%
%
\begin{table*}{{\bf Table \ref{results}} - cont.}
\begin{center}
\begin{tabular}{ccccccccccccccc}
Cluster & $\dot{M}$ & $r_{\rm cool}$ & $t_{\rm cool}$ & $L(<{r}_{\rm cool})$ & 
$\frac{{L}(<{r}_{\rm cool})}{{L}_{\rm bol}}$ & ${t}_{250}$ &
$M_{\rm gas}$ & $M_{\rm grav}$ & $\frac{{M}_{\rm gas}}{{M}_{\rm grav}}$ \\
\hline
(a) & (b) & (c) & (d) & (e) & (f) & (g) & (h) & (i) & (j) \\
\\ A754(P) & 0$^{+29}_{-0}$ & 0$^{+97}_{-0}$ & 15.0$^{+3.1}_{-2.2}$ & 0$^{+0.7}_{-0}$ & 
0$^{+0.04}_{-0}$ & 46.7$^{+6.8}_{-3.9}$ & 24.3$\pm$0.5 & 189 & 13$\pm$0.3  \\
\\ A754(H) & 2$^{+5}_{-2}$ & 26$^{+31}_{-26}$ & 7.9$^{+9.7}_{-2.1}$ & 0$^{+0}_{-0}$ & 
0$^{+0}_{-0}$ & 36.6$^{+7.4}_{-5.4}$ & 5.9$\pm$0.2 & 48 & 12$\pm$0.4  \\
\\ HYD-A(H) & $>$298 & $>$167 & 1.1$^{+0.1}_{-0.1}$ & $>$4.7 & 
$>$0.51 & $----$ & $----$ & $----$ & $----$  \\
\\ HYD-A(P) & 264$^{+81}_{-60}$ & 162$^{+56}_{-68}$ & 2.0$^{+0.0}_{-0.0}$ & 4.7$^{+1.0}_{-1.2}$ & 
0.52$^{+0.11}_{-0.13}$ & 21.0$^{+1.1}_{-1.0}$ & 26.5$\pm$0.3 & 166 & 16$\pm$0.2  \\
\\ A1060(P) & 15$^{+5}_{-7}$ & 79$^{+15}_{-26}$ & 4.7$^{+0.4}_{-0.3}$ & 0.15$^{+0.04}_{-0.1}$ & 
0.23$^{+0.06}_{0.15}$ & 39.9$^{+3.7}_{-2.9}$ & 2.8$\pm$0.0$^{\dagger}$ & 40$^{\dagger}$ & 7$^{\dagger}$   \\
\\ A1060(H) & 8$^{+3}_{-2}$ & 64$^{+9}_{-13}$ & 3.2$^{+1.2}_{-0.8}$ & 0.1$^{+0.01}_{-0.01}$ & 
15$^{+1.5}_{-1.5}$ & $----$ & $----$ &  $----$ \\
\\ A1367(P) & 0$^{+1}_{-0}$ & 0$^{+23}_{-0}$ & 25.8$^{+13.1}_{-8.0}$ & 0$^{}_{}$ & 
0 & 44.1$^{+3.9}_{-3.4}$ & 2.5$\pm$0.0$^{\dagger}$ & 23$^{\dagger}$ & 11$^{\dagger}$  \\
\\ VIRGO(H) & $>$12 & $>$28 & 0.1$^{+0.0}_{-0.0}$ & $>$0.1 & 
$>$0.37 & $----$ & $----$ & $----$ & $----$  \\
\\ VIRGO(P) & 39$^{+2}_{-9}$ & 102$^{+6}_{-4}$ & 0.2$^{+0.0}_{-0.0}$ & 0.3$^{+0.01}_{-0.01}$ & 
$\sim$0.5 & $----$ & $----$ & $----$ & $----$  \\
\\ CENT(H) & $>$24 & $>$67 & 0.4$^{+0.0}_{-0.0}$ & $>$0.4 & 
0.41$_{>0.1}$ & $----$ & $----$ & $----$ & $----$ \\
\\ CENT(P) & 30$^{+10}_{-5}$ & 81$^{+23}_{-18}$ & 0.8$^{+0.0}_{-0.0}$ & 0.5$^{+0.1}_{-0.1}$ & 
0.41$^{+0.1}_{-0.1}$ & 43.2$^{+14.3}_{-8.1}$ & 3.2$\pm$0.1$^{\dagger}$ & 34$^{\dagger}$ & 9$\pm$0.3$^{\dagger}$  \\
\\ COMA(P) & 0$^{+1}_{-0}$ & 0$^{+15}_{-0}$ & 17.7$^{+6.7}_{-4.1}$ & 0$^{+0}_{-0}$ & 
0 & 35.5$^{+2.0}_{-1.8}$ & 27.8$\pm$0.2 & 220 & 13$\pm$0.1 \\
\\ COMA(H) & 0$^{+2}_{-0}$ & 20$^{+19}_{-20}$ & 7.3$^{+5.2}_{-1.9}$ & 0.01$^{+0}_{-0}$ & 
0 & $----$ & $----$ & $----$ & $----$ \\
\\ A1644(H) & 11$^{+40}_{-5}$ & 58$^{+56}_{-20}$ & 2.2$^{+0.7}_{-0.6}$ & 0.2$^{+0.2}_{-0.1}$ & 
0.04$^{+0.04}_{-0.02}$ & $----$ & $----$ & $----$ & $----$  \\
\\ A3532(P) & 0$^{+25}_{-0}$ & 0$^{+104}_{-0}$ & 14.0$^{+3.3}_{-2.3}$ & 0$^{+0.3}_{-0}$ & 
0$^{+0.08}_{-0}$ & 40.8$^{+8.2}_{-5.9}$ & 18.2$\pm$0.6 & 132 & 14$\pm$0.5 \\
\\ A1650(H) & 280$^{+464}_{-89}$ & 165$^{+103}_{-24}$ & 2.4$^{+1.2}_{-0.8}$ & 5.4$^{+4.1}_{-0.9}$ & 
0.34$^{+0.26}_{-0.06}$ & 13.0$^{+3.7}_{-3.3}$ & 10.1$\pm$0.6$^{\dagger}$ & 60$^{\dagger}$ & 17$\pm$1$^{\dagger}$  \\
\\ A1651(P) & 138$^{+48}_{-41}$ & 127$^{+32}_{-31}$ & 6.5$^{+0.7}_{-0.7}$ & 3.3$^{+1.0}_{-1.0}$ & 
0.14$^{+0.04}_{-0.04}$ & 25.4$^{+2.2}_{-2.0}$ & 29.9$\pm$0.6 & 250 & 12$\pm$0.2  \\
\\ A1689(P) & 645$^{+196}_{-42}$ & 191$^{+103}_{-13}$ & 5.7$^{+0.2}_{-0.2}$ & 26.1$^{+9.5}_{-1.5}$ & 
0.43$^{+0.16}_{-0.02}$ & 22.8$^{+1.2}_{-0.9}$ & 64.6$\pm$0.9 & 476 & 14$\pm$0.2 \\
\\ A1689(H) & 484$^{+275}_{-93}$ & 162$^{+73}_{-21}$ & 2.5$^{+0.8}_{-0.5}$ & 18.0$^{+1.0}_{-1.0}$ & 
0.30$^{+0.02}_{-0.02}$ & 23.7$^{+18.9}_{-6.2}$ & 57.7$\pm$3.2 & 420 & 14$\pm$0.8 \\
\\ A1736(H) & 1$^{+4}_{-1}$ & 24$^{+50}_{-24}$ & 7.9$^{+15.3}_{-4.6}$ & 0.02$^{+0.03}_{-0}$ & 
0.01$^{+0.01}_{-0}$ & 47.8$^{+63.9}_{-18.1}$ & 2.8$\pm$0.3$^{\dagger}$ & 21$^{\dagger}$ & 13$\pm$1.4$^{\dagger}$ \\
\\ A3558(H) & 40$^{+21}_{-31}$ & 90$^{+24}_{-52}$ & 2.4$^{+2.9}_{-1.6}$ & 0.8$^{+0.4}_{-0.6}$ & 
0.08$^{+0.04}_{-0.16}$ & 41.4$^{+14.4}_{-6.1}$ & 5.8$\pm$0.2$^{\dagger}$ & 56$^{\dagger}$ & 10$\pm$0.4$^{\dagger}$ \\
\\ A3558(P) & 40$^{+39}_{-10}$ & 68$^{+75}_{-20}$ & 10.2$^{+0.3}_{-0.2}$ & 1.0$^{+0.9}_{-0.2}$ & 
0.1$^{+0.09}_{-0.02}$ & 35.9$^{+1.1}_{-1.1}$ & 31.5$\pm$0.2 & 226 & 14$\pm$0.1  \\
\\ A3562(P) & 37$^{+26}_{-27}$ & 95$^{+55}_{-65}$ & 7.2$^{+0.6}_{-0.5}$ & 0.5$^{+0.3}_{-0.3}$ & 
0.05$^{+0.03}_{-0.03}$ & 38.8$^{+4.3}_{-3.0}$ & 17.3$\pm$0.3 & 105 & 16$\pm$0.3  \\
\\ A3571(H) & 81$^{+32}_{-34}$ & 117$^{+23}_{-27}$ & 4.1$^{+0.7}_{-0.6}$ & 1.8$^{+0.6}_{-0.7}$ & 
0.1$^{+0.03}_{-0.04}$ & 26.4$^{+3.0}_{-3.3}$ & 7.8$\pm$0.2$^{\dagger}$ & 81$^{\dagger}$ & 10$\pm$0.3$^{\dagger}$  \\
\\ A3571(P) & 72$^{+56}_{-31}$ & 104$^{+39}_{-24}$ & 5.8$^{+0.9}_{-1.0}$ & 1.6$^{+1.2}_{-0.5}$ & 
0.09$^{+0.07}_{-0.03}$ & 27.0$^{+3.1}_{-2.1}$ & 30.0$\pm$0.5 & 220 & 14$\pm$0.2  \\
\\ A1795(P) & 381$^{+41}_{-23}$ & 177$^{+19}_{-6}$ & 1.9$^{+0.1}_{-0.1}$ &  8.5$^{+0.6}_{-0.2}$ & 
0.44$^{+0.03}_{-0.01}$ & 20.7$^{+1.4}_{-1.5}$ & 31.9$\pm$0.3 & 133 & 23$\pm$0.2  \\
\\ A1795(H) & 488$^{+274}_{-166}$ & 205$^{+77}_{-54}$ & 0.8$^{+0.2}_{-0.1}$ &  10.4$^{+0.4}_{-0.4}$ & 
0.54$^{+0.0}_{-0.0}$ & 16.4$^{+8.0}_{-4.2}$ & 11.4$\pm$0.4 & 79 & 14$\pm$0.5  \\
\\ A2029(H) & 554$^{+215}_{-93}$ & 179$^{+47}_{-13}$ & 1.0$^{+0.1}_{-0.1}$ & 16.8$^{+3.5}_{-1.1}$ & 
0.39$^{+0.08}_{-0.03}$ & 15.3$^{+3.5}_{-2.6}$ & 13.4$\pm$0.3$^{\dagger}$ & 116$^{\dagger}$ & 12$\pm$0.3$^{\dagger}$ \\
\\ A2029(P) & 556$^{+44}_{-73}$ & 186$^{+19}_{-39}$ & 2.9$^{+0.1}_{-0.1}$ & 17.8$^{+1.2}_{-2.9}$ & 
0.42$^{+0.03}_{-0.07}$ & 21.5$^{+1.0}_{-1.1}$ & 44.5$\pm$0.5 & 251 & 18$\pm$0.2  \\
\\ A2052(P) & 125$^{+26}_{-6}$ & 147$^{+53}_{-3}$ & 2.5$^{+0.1}_{-0.1}$ & 1.9$^{+0.4}_{-0.02}$ & 
0.51$^{+0.11}_{-0.01}$ & 26.0$^{+1.9}_{-2.0}$ & 15.9$\pm$0.4 & 107 & 14$\pm$0.4  \\
\\ A2052(H) & 102$^{+108}_{-15}$ & 134$^{+71}_{-20}$ & 1.3$^{+0.3}_{-0.2}$ & 1.7$^{+0.2}_{-0.1}$ & 
0.46$^{+0.5}_{-0.3}$ & 32.0$^{+45.1}_{-7.7}$ & 5.1$\pm$0.3 & 49 & 10$\pm$0.6  \\
\end{tabular}
\end{center}
\end{table*}
%
%
\begin{table*}{{\bf Table \ref{results}} - cont.}
\begin{center}
\begin{tabular}{ccccccccccccccc}
Cluster & $\dot{M}$ & $r_{\rm cool}$ & $t_{\rm cool}$ & $L(<{r}_{\rm cool})$ & 
$\frac{{L}(<{r}_{\rm cool})}{{L}_{\rm bol}}$ & ${t}_{250}$ &
$M_{\rm gas}$ & $M_{\rm grav}$ & $\frac{{M}_{\rm gas}}{{M}_{\rm grav}}$ \\
\hline
(a) & (b) & (c) & (d) & (e) & (f) & (g) & (h) & (i) & (j)  \\
\\ MKW3(P) & 175$^{+14}_{-46}$ & 171$^{+9}_{-62}$ & 3.0$^{+0.1}_{-0.1}$ & 2.3$^{+0.1}_{-0.6}$ & 
0.52$^{+0.02}_{-0.14}$ & 22.7$^{+1.8}_{-1.6}$ & 7.1$\pm$0.1$^{\dagger}$ & 58$^{\dagger}$ & 12$\pm$0.2$^{\dagger}$  \\
\\ MKW3(H) & 107$^{+150}_{-50}$ & 137$^{+51}_{-7}$ & 1.1$^{+0.9}_{-0.3}$ & 1.4$^{+0.4}_{-0.5}$ & 
0.32$^{+0.09}_{-0.11}$ & $----$ & $----$ & $----$ & $----$  \\
\\ A2065(H) & 13$^{+14}_{-6}$ & 56$^{+22}_{-23}$ & 4.4$^{+2.2}_{-1.3}$ &  0.5$^{+0.2}_{-0.2}$ & 
0.04$^{+0.01}_{-0.01}$ & 47.6$^{+50.6}_{-16.0}$ & 6.8$\pm$0.5$^{\dagger}$ & 90$^{\dagger}$ & 8$\pm$0.6$^{\dagger}$  \\
\\ A2063(P) & 37$^{+7}_{-12}$ & 95$^{+13}_{-30}$ & 5.0$^{+0.4}_{-0.3}$ & 0.6$^{+0.1}_{-0.3}$ & 
0.20$^{+0.03}_{-0.1}$ & 30.8$^{+2.6}_{-2.9}$ & 15.8$\pm$0.3 & 116 & 14$\pm$0.3  \\
\\ A2142(P) & 350$^{+66}_{-133}$ & 150$^{+18}_{-49}$ & 5.2$^{+0.4}_{-0.3}$ & 13.0$^{+1.8}_{-4.6}$ & 
0.24$^{+0.03}_{-0.09}$ & 17.2$^{+0.8}_{-0.8}$ & 52.0$\pm$0.8 & 369 & 14$\pm$0.2 \\
\\ A2142(H) & 286$^{+57}_{-74}$ & 152$^{+23}_{-31}$ & 3.8$^{+1.9}_{-0.8}$ & 11.2$^{+2.1}_{-2.8}$ & 
0.21$^{+0.02}_{-0.02}$ & 26.6$^{+6.8}_{-4.2}$ & 47.2$\pm$0.3 & 349 & 14$\pm$0.1 \\
\\ A2199(H) & 171$^{+76}_{-28}$ & 152$^{+17}_{-23}$ & 1.2$^{+0.2}_{-0.1}$ & 3.0$^{+0.2}_{-0.5}$ & 
0.47$^{+0.03}_{-0.08}$ & $----$ & $----$ & $----$ & $----$ \\
\\ A2199(P) & 154$^{+18}_{-8}$ & 143$^{+17}_{-6}$ & 1.9$^{+0.0}_{-0.1}$ & 2.7$^{+0.3}_{-0.1}$ & 
0.42$^{+0.05}_{-0.02}$ & 27.9$^{+1.3}_{-1.5}$ & 20.8$\pm$0.2 & 146 & 14$\pm$0.1 \\
\\ A2204(P) & 852$^{+127}_{-82}$ & 199$^{+60}_{-44}$ & 3.1$^{+0.1}_{-0.1}$ & 40.1$^{+4.6}_{-4.6}$ & 
0.75$^{+0.09}_{-0.09}$ & 19.4$^{+1.2}_{-1.6}$ & 51.6$\pm$1.2 & 339 & 15$\pm$0.4 \\
\\ A2204(H) & 843$^{+245}_{-152}$ & 181$^{+88}_{-36}$ & 1.0$^{+0.1}_{-0}$ & 40.3$^{+8.8}_{-4.2}$ & 
0.75$^{+0.16}_{-0.08}$ & 23.8$^{+7.4}_{-4.5}$ & 21.0$\pm$0.6$^{\dagger}$ & 186$^{\dagger}$ & 11$\pm$0.3$^{\dagger}$  \\
\\ TRI AUST(P) & 33$^{+87}_{-33}$ & 76$^{+67}_{-76}$ & 10.8$^{+7.1}_{-3.1}$ & 0.7$^{+1.7}_{-0}$ & 
0.03$^{+0.06}_{-0}$ & 24.4$^{+1.9}_{-2.3}$ & 38.8$\pm$0.7 & 197 & 20$\pm$0.4 \\
\\ A2244(P) & 244$^{+49}_{-145}$ & 148$^{+20}_{-92}$ & 7.3$^{+0.6}_{-0.5}$ & 6.6$^{+0.8}_{-4.0}$ & 
0.43$^{+0.05}_{-0.26}$ & 27.9$^{+3.2}_{-2.5}$ & 35.7$\pm$1.2 & 284 & 13$\pm$0.4  \\
\\ A2256(P) & 0$^{+14}_{-0}$ & 0$^{+69}_{-0}$ & 15.0$^{+4.0}_{-3.6}$ & 0$^{+0.2}_{-0}$ & 
0$^{+0.01}_{-0}$ & 35.5$^{+3.9}_{-3.6}$ & 24.4$\pm$0.4 & 154 & 16$\pm$0.3 \\
\\ A2256(H) & 0$^{+16}_{-0}$ & 16$^{+69}_{-16}$ & 10.8$^{+20.5}_{-9.1}$ & 0.0$^{+0.26}_{-0}$ & 
0$^{+0.0}_{-0.0}$ & 28.0$^{+5.2}_{-3.9}$ & 26.5$\pm$0.3 & 159 & 17$\pm$0.3 \\
\\ OPHI(P) & 127$^{+48}_{-94}$ & 129$^{+22}_{-70}$ & 3.0$^{+0.4}_{-0.3}$ & 3.8$^{+0.9}_{-2.6}$ & 
0.12$^{+0.03}_{-0.08}$ & 20.1$^{+1.9}_{-1.8}$ & 37.6$\pm$0.7 & 242 & 16$\pm$0.3  \\
\\ OPHI(H) & 34$^{+10}_{-11}$ & 72$^{+7}_{-1}$ & 1.0$^{+0.1}_{-0.1}$ & 1.3$^{+0.1}_{-0.01}$ & 
0.04 & $----$ & $----$ & $----$ & $----$  \\
\\ A2255(P) & 0$^{+4}_{-0}$ & 0$^{+46}_{-0}$ & 36.9$^{+21.0}_{-9.5}$ & 0$^{+0.1}_{-0}$ & 
0$^{+0.01}_{-0}$ & 51.2$^{+7.0}_{-6.9}$ & 21.5$\pm$0.5 & 171 & 13$\pm$0.3 \\
\\ A2319(H) & 4$^{+103}_{-4}$ & 34$^{+100}_{-34}$ & 9.4$^{+15.5}_{-3.7}$ & 0.1$^{+2.6}_{-0}$ & 
0$^{+0.07}_{-0}$ & 31.6$^{+16.6}_{-8.6}$  & 9.9$\pm$0.5$^{\dagger}$ & 103$^{\dagger}$ & 10$\pm$0.5$^{\dagger}$   \\
\\ A2319(P) & 20$^{+61}_{-20}$ & 53$^{+59}_{-53}$ & 10.8$^{+6.0}_{-2.9}$ & 0.5$^{+1.5}_{-0}$ & 
0.01$^{+0.04}_{-0}$ & 26.6$^{+4.0}_{-3.0}$ & 37.0$\pm$1.2 & 240 & 15$\pm$0.5 \\
\\ CYG-A(H) & 215$^{+26}_{-22}$ & 121$^{+11}_{-12}$ & 1.7$^{+0.1}_{-0.1}$ & 8.1$^{+0.3}_{-0.4}$ & 
0.49$^{+0.02}_{-0.02}$ & 28.5$^{+9.4}_{-6.6}$ & 9.0$\pm$0.3$^{\dagger}$ & 109$^{\dagger}$ & 8$\pm$0.3$^{\dagger}$ \\
\\ CYG-A(P) & 244$^{+26}_{-22}$ & 135$^{+33}_{-34}$ & 2.6$^{+0.1}_{-0.1}$ & 9.3$^{+0.8}_{-1.1}$ & 
0.56$^{+0.05}_{-0.07}$ & 28.5$^{+3.6}_{-2.1}$ & 28.2$\pm$0.8 & 201 & 14$\pm$0.4 \\
\\ A3667(P) & 0$^{+11}_{-0}$ & 0$^{+63}_{-0}$ & 13.4$^{+6.4}_{-3.4}$ & 0$^{+0.2}_{-0}$ & 
0$^{+0.01}_{-0}$ & 28.7$^{+2.9}_{-2.3}$ & 24.8$\pm$0.4 & 153 & 16$\pm$0.3  \\
\\ A2597(H) & 276$^{+60}_{-3}$ & 139$^{+24}_{-26}$ & 1.4$^{+0.1}_{-0.1}$ & 8.2$^{+0.7}_{-0.9}$ & 
0.61$^{+0.05}_{-0.07}$ & $----$ & $----$ & $----$ & $----$ \\
\\ A2597(P) & 271$^{+41}_{-41}$ & 152$^{+67}_{-58}$ & 2.3$^{+0.1}_{-0.1}$ & 8.2$^{+1.3}_{-1.9}$ & 
0.61$^{+0.1}_{-0.14}$ & 32.7$^{+5.3}_{-2.9}$ & 22.9$\pm$0.6 & 217 & 11$\pm$0.3 \\
\\ KLEM44(P) & 87$^{+25}_{-19}$ & 133$^{+42}_{-27}$ & 2.3$^{+0.6}_{-0.3}$ & 1.1$^{+0.3}_{-0.2}$ & 
0.49$^{+0.13}_{-0.09}$ & 83.8$^{+121.0}_{-42.7}$ & 4.3$\pm$0.2$^{\dagger}$ & 50$^{\dagger}$ & 9$\pm$0.4$^{\dagger}$ \\
\\ A4059(H) & 110$^{+72}_{-33}$ & 143$^{+34}_{-12}$ & 1.9$^{+1.14}_{-0.6}$ & 1.5$^{+0.4}_{-0.1}$ & 
0.32$^{+0.09}_{-0.02}$ & $----$ & $----$ & $----$ & $----$ \\
\\ A4059(P) & 130$^{+27}_{-21}$ & 153$^{+19}_{-18}$ & 3.4$^{+0.6}_{-0.4}$ & 1.6$^{+0.2}_{-0.2}$ & 
0.34$^{+0.04}_{-0.04}$ & 24.8$^{+3.6}_{-3.6}$ & 19.6$\pm$0.6 & 98 & 20$\pm$0.6  \\
\end{tabular}
\end{center}
\vskip 1.0 cm
\begin{flushleft}
{\bf Notes.} The errors quoted for $r_{\rm cool}$ are smaller than the resolution used. 
This must be understood 
remembering that they are uncertainties computed from the deprojection of Monte-Carlo perturbations 
of the X-ray surface brightness. 
The limits in $L(<{r}_{\rm cool})$ correspond to the bolometric luminosities within the 
extrema of $r_{\rm cool}$.
We use the integrated X-ray luminosity at the last deprojected bin as the X-ray bolometric luminosity 
for A3532, since the latter is not available in David et al. (1992). The luminosity for M87 in the Virgo
cluster is taken from White et al. (1997).
 
\end{flushleft}
\end{table*}
\end{document}